\documentclass[fleqn,12pt]{article}

\usepackage{graphics,epsfig}

\newcommand{\be}[1]{\begin{equation} \label{(#1)}}
\newcommand{\ee}{\end{equation}}
\newcommand{\baq}[1]{\begin{eqnarray} \label{(#1)}}
\newcommand{\eaq}{\end{eqnarray}}

\newcommand{\rf}[1]{(\ref{(#1)})}

\def\lsim{\raise0.3ex\hbox{$\;<$\kern-0.75em\raise-1.1ex\hbox{$\sim\;$}}}
\def\gsim{\raise0.3ex\hbox{$\;>$\kern-0.75em\raise-1.1ex\hbox{$\sim\;$}}}

\def\beq   {\begin{equation}}
\def\eeq   {\end{equation}}
\def\beqd  {\begin{displaymath}}
\def\eeqd  {\end{displaymath}}
\def\beqaa {\begin{eqnarray}}
\def\eeqaa {\end{eqnarray}}

\def\ti  {\tilde}

\def\su  {\ti u}

\def\sc  {\ti c}

\def\st  {\ti t}

\def\sg  {\ti g}

\def\nt  {\tilde\chi^0}
\def\ch  {\tilde\chi^\pm}
\def\chp {\tilde\chi^+}

\def\a   {\alpha}
\def\b   {\beta}
\def\t   {\theta}

\def\gev {\rm ~GeV}

\def\sz{\ifmmode{\tilde{\chi}^0} \else{$\tilde{\chi}^0$} \fi}
\def\sw{\ifmmode{\tilde{\chi}} \else{$\tilde{\chi}$} \fi}

\newcommand{\AddrVienna}{
\it  Fakult\"at f\"ur Physik, Universit\"at Wien, 
A-1090 Vienna, Austria \\}

\newcommand{\AddrGAKUGEI}{%
 \it Department of Physics, Tokyo Gakugei University, Koganei, 
Tokyo 184-8501, Japan\\}

\newcommand{\AddrHEPHY}{%
 \it Institut f\"ur Hochenergiephysik der \"Osterreichischen Akademie 
der Wissenschaften, A-1050 Vienna, Austria\\}

\newcommand{\AddrWuerzburg}{%
 \it Institut f\"ur Theoretische Physik und Astrophysik, Universit\"at W\"urzburg, 
D-97074 W\"urzburg, Germany\\}

\newcommand{\AddrAHEP}{
\it AHEP Group, Instituto de Fisica Corpuscular - C.S.I.C., Universidad de Val{\`e}ncia, \\ 
E-46071 Val{\`e}ncia, Spain \\}

\newcommand{\AddrDESY}{
\it  Deutsches Elektronen-Synchrotron (DESY), Theory Group, 
D-22603 Hamburg, Germany \\}

\begin{document}

\begin{flushright}
DESY 10-059
\end{flushright}

\begin{center}  
  \textbf{\Large 
Impact of squark generation mixing on the search for squarks decaying into 
fermions at LHC}\\[10mm]
{A. Bartl${}^{1}$, H. Eberl${}^2$, B. Herrmann${}^{3,4}$, K.~Hidaka${}^5$, 
W.~Majerotto${}^2$ and W. Porod${}^{4,6}$
} 
\vspace{0.1cm}\\ 
$^1$ \AddrVienna
$^2$ \AddrHEPHY
$^3$ \AddrDESY
$^4$ \AddrWuerzburg
$^5$ \AddrGAKUGEI
$^6$ \AddrAHEP
%
%
\end{center}

\noindent

\begin{abstract}
\noindent
We study the effect of squark generation mixing on squark production and decays 
at LHC in the Minimal Supersymmetric Standard Model (MSSM). We show that the effect 
can be very large despite the very 
strong constraints on quark-flavour violation (QFV) from experimental data on B mesons. 
We find that the two lightest up-type squarks $\su_{1,2}$ can have large 
branching ratios for the decays into $c \tilde \chi^0_1$ and 
$t \tilde \chi^0_1$ at the same time due to squark generation mixing, leading to QFV 
signals '$pp \to c \bar{t} (t \bar{c})$ + missing-$E_T$ + $X$' with a significant rate. 
The observation of this remarkable signature would provide a powerful test of 
supersymmetric QFV at LHC. This could have a significant impact on the search 
for squarks and the determination of the underlying MSSM parameters.
%

%
\end{abstract}

\newpage 

\section{Introduction}

The exploration of the TeV scale has begun with the start up of the LHC run. 
Gluinos and squarks, the supersymmetric partners of gluons and quarks, 
will be produced copiously for masses up to $O(1 ~TeV)$ if supersymmetry 
(SUSY) is realized in nature. After the discovery of SUSY, the determination 
of SUSY parameters will be one of the main experimental programs. 
The determination of the soft-SUSY-breaking parameters will be particularly 
important to pin down the SUSY-breaking mechanism. As the soft-SUSY-breaking 
terms are the source of flavour violation beyond the Standard Model (SM), 
the measurement of flavour violating observables is directly linked to the 
crucial question about the SUSY-breaking mechanism. 
It is usually assumed that production and decays of gluinos and squarks are 
quark-flavour conserving (QFC). However, additional flavour structures 
(i.e. squark generation mixings) would imply that squarks are not quark-flavour 
eigenstates, which could result in sizable quark-flavour violation (QFV) effects 
significantly larger than those due to the Cabibbo-Kobayashi-Maskawa (CKM) mixing. 
\\
Additional flavour structures will of course give contributions to
flavour-changing neutral current (FCNC) processes.  Up to now all
measurements of such processes are consistent with the SM 
predictions, which in turn requires that the flavour structure of new
physics at the TeV scale is highly constrained. In particular, this 
flavour structure could be closely related to the flavour
structure of the SM Yukawa couplings. 
The most extreme case is minimal flavour violation (MFV) 
\cite{Buras:2000dm,Ambrosio:2002ex,Kagan:2009bn} which assumes that 
the Yukawa coupling matrices of the SM are the only source of
flavour violation even in interactions involving new particles.
Supersymmetric models of this kind are gauge-mediated SUSY-breaking 
or minimal supergravity (mSUGRA) models with universal boundary 
conditions \cite{Colangelo:2008qp}.
However, while the flavour constraints suggest that the dominant flavour
structure of new physics should be MFV, there is certainly room for
sub-dominant contributions that are not MFV. The discovery of such
non-MFV (NMFV) physics will be of utmost interest.
There are also known examples of flavour models which do have large
flavour violating entries in the squark sector getting consistency
with the flavour observables in a different way. An example is a model with 
an extended R-symmetry \cite{Kribs:2007ac} where the left-right squark mixing
terms are absent and the gauginos are Dirac particles. Another possibility
would be hybrid gauge and gravity mediation of supersymmetry 
breaking \cite{ArkaniHamed:1997km} where one gets sizable NMFV contributions 
(i.e. sizable squark generation mixing terms) as discussed in \cite{Hiller:2008sv}.
\\
The effect of QFV in the squark sector on reactions with external particles 
being SM particles \cite{QFV_quark_pair@LHC, QFV_top_decay} 
(or SUSY Higgs bosons \cite{QFV_Higgs_decay}) has been studied in several 
publications. In this case the effect of QFV in the squark sector is induced 
only by SUSY particle (sparticle) loops. \\
However, in reactions with external SUSY particles, the QFV effect can 
already occur at tree-level and hence can be rather large. 
The QFV decay $\tilde t_1\to c\tilde \chi_1^0$ \cite{QFV_decay_in_MFV} and 
QFV gluino decays \cite{Porod:2002wz} were studied in the scenario of MFV, 
where the  only source of QFV is the mixing due to 
the CKM matrix. The QFV decay $\tilde t_1\to c\tilde \chi_1^0$ is actually the 
standard search mode at the Tevatron for light top-squarks if their decays into 
bottom-quark plus chargino and top-quark plus neutralino are kinematically 
forbidden. Squark pair production and their decays at LHC have been analyzed 
in scenarios of NMFV, where the effect of the squark generation mixing is also 
included \cite{Bozzi:2007me, QFV_squark_decay}. QFV gluino decays 
\cite{Bartl:QFV_gluino_decay} and QFV squark decays \cite{Hurth_Porod} have been 
studied in the Minimal Supersymmetric Standard Model (MSSM) with squark 
generation mixing in its most general form. 

In the present paper, we study the effect of QFV due to the mixing of 
charm-squarks and top-squarks both on production and subsequent decays of 
squarks in the general MSSM with R parity conservation. 
In principle also the mixing between right up-squark and left top-squark is 
hardly constrained as pointed out in \cite{1-3 mixing}. Here for simplicity we 
do not take into account such a mixing as we are mainly interested in 
demonstrating the main QFV effects and signals. Note that in case one cannot 
distinguish between the quarks of the first two generations, the corresponding 
QFV signals will involve jets whose original quark is not identified, and hence 
the effects of the two mixings (i.e. the 1st and 3rd generation mixing and the 
2nd and  3rd generation mixing) cannot be distinguished. \\
%
We show that the QFV squark decay branching ratios B($\su_i \to c \nt_1$) and 
B($\su_i \to t \nt_1$) $(i=1,2)$ can be very large (up to $\sim$ 50\%) 
simultaneously due to the squark generation mixing in a significant region of 
the QFV parameters despite the very strong experimental constraints 
from B factories, Tevatron and LEP.
Here $\su_{1,2}$ are the two lightest up-type squarks and $\nt_1$ is the 
lightest neutralino. This leads to QFV signal events 
'$pp \to$ $c$${\bar t}$ (${\bar c}$$t$) + $E_T^{mis}$ + $X$' and 
'$pp \to$ $t \, t$ (${\bar t} \, {\bar t}$) + $E_T^{mis}$ + $X$' at LHC, 
which we also study in the present article, where $E_T^{mis}$ is 
the missing transverse energy. 
%

\section{Squark mixing with flavour violation}
%

The most general up-type squark mass matrix including left-right mixing
as well as quark-flavour mixing in the super-CKM basis of 
$\tilde u_{0\gamma}=
(\tilde u_L,\tilde c_L,\tilde t_L,\tilde u_R,\tilde c_R,\tilde t_R)$, 
$\gamma=1,\dots,6$, is \cite{Allanach:2008qq}
%
%

\baq{eq:SquarkMassMatrix}
M^2_{\tilde u}=\left(\begin{array}{ccc}
M^2_{\tilde u LL} & (M^2_{\tilde u RL})^\dagger\\[5mm]
M^2_{\tilde u RL} & M^2_{\tilde u RR}
\end{array}\right)~,
\eaq
where the three $3\times3$ matrices read
\begin{eqnarray}
(M^2_{\tilde u LL})_{\alpha\beta} & = & 
M^2_{Q_u \alpha\beta}+\left[(\frac{1}{2}-\frac{2}{3}\sin^2\theta_W)
\cos2\beta~m_Z^2+m_{u_\alpha}^2\right]\delta_{\alpha\beta},
\label{eq:LL}\\[3mm]
(M^2_{\tilde u RR})_{\alpha\beta} & = & M_{U\alpha\beta}^2
+\left[\frac{2}{3}\sin^2\theta_W\cos2\beta~
m_Z^2+m_{u_\alpha}^2\right] \delta_{\alpha\beta}~,
\label{eq:RR}\\[3mm]
(M^2_{\tilde u RL})_{\alpha\beta} & = & (v_2/\sqrt{2} ) T_{U\beta\alpha}-
m_{u_\alpha} \mu^*\cot\beta~\delta_{\alpha\beta}~.
\label{eq:RL}
\end{eqnarray}
The indices $\alpha,\beta=1,2,3$ characterize the quark flavours $u,c,t$, respectively.
$M_{Q_u}^2$ and $M_U^2$ are the hermitian soft-SUSY-breaking mass matrices for the left 
and right up-type squarks, respectively. Note that in the super-CKM basis one has 
$M_{Q_u}^2 = K\cdot M_Q^2\cdot K^\dagger $ due to the SU(2) symmetry, where $M_Q^2$ is 
the hermitian soft-SUSY-breaking mass matrix for the left down-type squarks and 
$K$ is the CKM matrix. 
Note also that $M_{Q_u}^2 \simeq M_Q^2$ as $K \simeq 1$. 
$T_U$ is the soft-SUSY-breaking trilinear coupling matrix of the up-type squarks: 
${\mathcal L}_{\rm int}=
-(T_{U\alpha\beta} \tilde u^\dagger_{R\beta}\tilde  u_{L\alpha} H^0_2 + h.c.) + \cdots$.
$\mu$ is the higgsino mass parameter. 
$v_{1,2}$ are the vacuum expectation values of the Higgs fields with 
$v_{1,2}/\sqrt{2} \equiv \langle H^0_{1,2}\rangle$, and $\tan\beta \equiv v_2/v_1$. 
%
%
%
$m_{u_\a}$ $(u_\a=u,c,t)$ are the quark masses.\\
The physical mass eigenstates $\tilde u_i$, $i=1,\dots,6$, are given
by $\tilde u_i=R^{\tilde u}_{i\alpha}\tilde u_{0\alpha}$. 
The mixing matrix $R^{\tilde u}$ and the mass eigenvalues 
are obtained by a unitary transformation 
$R^{\tilde u} M^2_{\tilde u} R^{\tilde u\dagger}=
{\rm diag}(m_{\tilde u_1},\dots,m_{\tilde u_6})$, where $m_{\tilde u_i}<m_{\tilde u_j}$
for $i<j$. \\
Having in mind that $M_{Q_u}^2 \simeq M_Q^2$, we define the QFV parameters 
$\delta^{uLL}_{\alpha\beta}$, $\delta^{uRR}_{\alpha\beta}$ 
and $\delta^{uRL}_{\alpha\beta}$ $(\alpha \neq \beta)$ as follows \cite{Gabbiani}: 
\begin{eqnarray}
\delta^{uLL}_{\alpha\beta} & \equiv & M^2_{Q \alpha\beta} / \sqrt{M^2_{Q \alpha\alpha} M^2_{Q \beta\beta}}~,
\label{eq:InsLL}\\[3mm]
\delta^{uRR}_{\alpha\beta} &\equiv& M^2_{U \alpha\beta} / \sqrt{M^2_{U \alpha\alpha} M^2_{U \beta\beta}}~,
\label{eq:InsRR}\\[3mm]
\delta^{uRL}_{\alpha\beta} &\equiv& (v_2/\sqrt{2} ) T_{U\beta\alpha} / \sqrt{M^2_{U \alpha\alpha} M^2_{Q \beta\beta}}~.
\label{eq:InsRL}
\end{eqnarray}
The relevant QFV parameters in this study are $\delta^{uLL}_{23}$, $\delta^{uRR}_{23}$, 
$\delta^{uRL}_{23}$ and $\delta^{uRL}_{32}$ which are the $\sc_L - \st_L$, $\sc_R-\st_R$, 
$\sc_R - \st_L$ and $\sc_L - \st_R$ mixing parameters, respectively. 
The down-type squark mass matrix can be parameterized analogously to the up-type 
squark mass matrix \cite{Allanach:2008qq}. 


The properties of the charginos $\ch_i$ ($i=1,2$, $m_{\ch_1}<m_{\ch_2}$) 
and neutralinos $\nt_k$ ($k=1,...,4$, $m_{\nt_1}< ...< m_{\nt_4}$)  
are determined by the parameters $M_2$, $M_1$, $\mu$ and $\tan\b$, 
where $M_2$ and $M_1$ are the SU(2) and U(1) gaugino mass parameters, respectively. 

\section{Constraints}\label{sec:constraints}

In our analysis, we impose the following conditions
on the MSSM parameter space in order to respect experimental
and theoretical constraints:

\renewcommand{\labelenumi}{(\roman{enumi})} 
\begin{enumerate}
\item Constraints from the B-physics experiments relevant mainly for
      the mixing between the second and third generations of squarks: \\
      $B(b \to s ~\gamma) = (3.57 \pm ((0.24 \times 1.96)^2 + (0.23 \times 1.96)^2)^{1/2}) 
      \times 10^{-4} = (3.57 \pm 0.65) \times 10^{-4}$ (95\% CL), 
      where we have combined the experimental error of $0.24 \times 1.96 \times 10^{-4}$ 
      (95\% CL) \cite{Iijima_LP2009} quadratically with the theoretical uncertainty
      of $0.23 \times 1.96 \times 10^{-4}$ (95\% CL) \cite{b_s_gamma_SM_error}, 
      $0.60 \times 10^{-6} <  B(b \to s ~ l^+l^-) < 2.60 \times 10^{-6}$ 
                                                  with $l=e ~{\rm or} ~\mu$ (95\% CL) 
                                                         \cite{b->sl+l-},
      $B(B_s \to \mu^+\mu^-) < 4.3 \times 10^{-8}$ (95\% CL) 
                                     \cite{Iijima_LP2009},
      $|R_{B\tau\nu}^{SUSY} - 1.35| < 0.76$ (95\% CL) 
      with $R_{B\tau\nu}^{SUSY} \equiv B^{SUSY}(B_u^- \to \tau^- {\bar\nu}_\tau) / 
      B^{SM}(B_u^- \to \tau^- {\bar\nu}_\tau) \simeq (1 - (\frac{m_{B^+}\tan\b}{m_{H^+}})^2)^2$ 
                                                         \cite{Sangro_SUSY2009}. 
      Moreover we impose the following condition on the SUSY prediction:
      $|\Delta M_{B_s}^{SUSY} - 17.77| < ((0.12 \times 1.96)^2 + 3.3^2)^{1/2} ~ps^{-1} 
      = 3.31 ~ps^{-1}$ (95\% CL),
      where we have combined the experimental error of $0.12 \times 1.96 ~ps^{-1}$
      (95\% CL) \cite{DMBs_CDF} quadratically with the theoretical uncertainty
      of $3.3 ps^{-1}$ (95\% CL) \cite{DMBs_theoretical_error}.
%
\item The experimental limit on SUSY contributions to the electroweak $\rho$ parameter \cite{rho_parameter}: 
$\Delta\rho(SUSY)<0.0012$.
\item The LEP limits on the SUSY particle masses 
        \cite{LEP}:
        $m_{\ch_1} > 103$ GeV, $m_{\nt_1} > 50$ GeV,
        $m_{\ti{u}_1,\ti{d}_1} > 100$ GeV, $m_{\ti{u}_1,\ti{d}_1} > m_{\nt_1}$,
        $m_{A^0} > 93~{\rm GeV}$, $m_{h^0}>110$~GeV, where $A^0$ is the CP-odd 
        Higgs boson and $h^0$ is the lighter CP-even Higgs boson. 
%
%
\item The Tevatron limits on the gluino and squark masses \cite{m_gluino_ICHEP2008}.
%
\item The vacuum stability conditions for the trilinear coupling matrix \cite{Casas}:
\begin{eqnarray}
|T_{U\alpha\alpha}|^2 &<&
3~Y^2_{U\alpha}~(M^2_{Q_u \alpha\alpha}+M^2_{U\alpha\alpha}+m^2_2)~,
\label{eq:CCBfcU}\\[2mm]
|T_{D\alpha\alpha}|^2 &<&
3~Y^2_{D\alpha}~(M^2_{Q\alpha\alpha}+M^2_{D\alpha\alpha}+m^2_1)~,
\label{eq:CCBfcD}\\[2mm]
|T_{U\alpha\beta}|^2 &<&
Y^2_{U\gamma}~(M^2_{Q_u \alpha\alpha}+M^2_{U\beta\beta}+m^2_2)~, 
\label{eq:CCBfvU}\\[2mm]
|T_{D\alpha\beta}|^2 &<&
Y^2_{D\gamma}~(M^2_{Q\alpha\alpha}+M^2_{D\beta\beta}+m^2_1)~,
\label{eq:CCBfvD}
\end{eqnarray}
with 
$(\a\neq\b;\gamma={\rm Max}(\a,\b);\a,\b=1,2,3)$ and 
$m^2_1=(m^2_{H^\pm}+m^2_Z\sin^2\theta_W)\sin^2\beta-\frac{1}{2}m_Z^2$,
$m^2_2=(m^2_{H^\pm}+m^2_Z\sin^2\theta_W)\cos^2\beta-\frac{1}{2}m_Z^2$.
The Yukawa couplings of the up-type and down-type quarks are 
$Y_{U\alpha}=\sqrt{2}m_{u_\alpha}/v_2=\frac{g}{\sqrt{2}}\frac{m_{u_\alpha}}{m_W \sin\beta}$ 
$(u_\a=u,c,t)$ and 
$Y_{D\alpha}=\sqrt{2}m_{d_\alpha}/v_1=\frac{g}{\sqrt{2}}\frac{m_{d_\alpha}}{m_W \cos\beta}$ 
$(d_\a=d,s,b)$, 
with $m_{u_\a}$ and $m_{d_\a}$ being the running quark masses at the weak scale and 
$g$ the SU(2) gauge coupling. All soft-SUSY-breaking parameters are assumed to be given 
at the weak scale. As SM parameters we take $m_W=80.4 \gev$, $m_Z=91.2 \gev$ and 
the on-shell top-quark mass $m_t=174.3 \gev$. 
We have found that our results shown in the following are fairly insensitive to the 
precise value of $m_t$.\\
\end{enumerate}
We calculate the observables in (i)-(iv) by using the public code SPheno v3.0 
\cite{SPheno_B-physics_refs}.
Condition (i) except for $B(B_u^- \to \tau^- {\bar\nu}_\tau)$ strongly constrains the 2nd and 3rd 
generation squark mixing parameters $M^2_{Q 23}, M^2_{U 23}, M^2_{D 23}, T_{U 23}, 
T_{D 23}$ and $T_{D32}$. The constraints from $B(b \to s \gamma)$ and 
$\Delta M_{B_s}$ are especially important \cite{Hurth_Porod}. $B(b \to s \gamma)$ 
is sensitive to $M^2_{Q 23}, T_{U 23}, T_{D 23}$ and $\Delta M_{B_s}$ is 
sensitive to $M^2_{Q 23} \cdot M^2_{U 23}$, $M^2_{Q 23} \cdot M^2_{D 23}$.\\
%
%

\section{Flavour violating fermionic squark decays}

%
%
We study the effect of the mixing between the 2nd and 3rd generation of squarks on 
their decays. The branching ratios of the squark decays 
\beq \label{eq:QFVsq_decay}
\su_{1,2} \to c ~ \nt_1 ~ \quad {\rm and} ~ \quad \su_{1,2} \to t ~ \nt_1
\eeq
are calculated by taking into account the following two--body decays: 
\baq{eq:decaymodes}
\tilde u_i &\to& u_k ~\sg, ~u_k ~\tilde\chi^0_n, ~d_k ~\tilde\chi^+_m, ~
\tilde{u}_j~Z^0, ~\tilde{d}_j~W^+, ~\tilde{u}_j~h^0,
\eaq
where $u_k=(u,c,t)$ and $d_k=(d,s,b)$.
The decays into the heavier Higgs bosons are kinematically 
forbidden in our scenarios studied below. 
The formulae for the widths of the two--body decays in \rf{eq:decaymodes} can be found in 
\cite{Bozzi:2007me}, except for the squark decays into the Higgs boson, for which 
we take the formulae of \cite{Bartl:2003pd, Bruhnke:2010rh}. 

\begin{table}[t]
\begin{center}

\begin{tabular}{|c||c|c|c|} \hline
 $M^2_{Q\alpha\beta}$
& \multicolumn{1}{c|}{\scriptsize{${\beta=1}$}} 
& \multicolumn{1}{c|}{\scriptsize{$\beta=2$}} 
& \multicolumn{1}{c|}{\scriptsize{$\beta=3$}} \\\hline\hline
 \scriptsize{$\alpha=1$}
& \multicolumn{1}{c|}{$(920)^2$} 
& \multicolumn{1}{c|}{0} 
& \multicolumn{1}{c|}{0} \\\hline

 \scriptsize{$\alpha=2$}
& \multicolumn{1}{c|}{0} 
& \multicolumn{1}{c|}{$(880)^2$} 
& \multicolumn{1}{c|}{$(224)^2$} \\\hline

 \scriptsize{$\alpha=3$}
& \multicolumn{1}{c|}{0} 
& \multicolumn{1}{c|}{$(224)^2$} 
& \multicolumn{1}{c|}{$(840)^2$} \\\hline
\end{tabular}
\begin{tabular}{|c|c|c|c|c|c|} \hline
 
  \multicolumn{1}{|c|}{$M_1$} 
& \multicolumn{1}{c|}{$M_2$} 
& \multicolumn{1}{c|}{$M_3$} 
& \multicolumn{1}{c|}{$\mu$} 
& \multicolumn{1}{c|}{$\tan\beta$} 
& \multicolumn{1}{c|}{$m_{A^0}$} \\\hline\hline
 
  \multicolumn{1}{|c|}{139} 
& \multicolumn{1}{c|}{264} 
& \multicolumn{1}{c|}{800} 
& \multicolumn{1}{c|}{1000} 
& \multicolumn{1}{c|}{10} 
& \multicolumn{1}{c|}{800} \\\hline

\end{tabular}
\vskip0.2cm
\begin{tabular}{|c||c|c|c|} \hline
 $M^2_{D\alpha\beta}$
& \multicolumn{1}{c|}{\scriptsize{${\beta=1}$}} 
& \multicolumn{1}{c|}{\scriptsize{$\beta=2$}} 
& \multicolumn{1}{c|}{\scriptsize{$\beta=3$}} \\\hline\hline
 \scriptsize{$\alpha=1$}
& \multicolumn{1}{c|}{$(830)^2$} 
& \multicolumn{1}{c|}{0} 
& \multicolumn{1}{c|}{0} \\\hline

 \scriptsize{$\alpha=2$}
& \multicolumn{1}{c|}{0} 
& \multicolumn{1}{c|}{$(820)^2$} 
& \multicolumn{1}{c|}{0} \\\hline

 \scriptsize{$\alpha=3$}
& \multicolumn{1}{c|}{0} 
& \multicolumn{1}{c|}{0} 
& \multicolumn{1}{c|}{$(810)^2$} \\\hline
\end{tabular}
\hspace{0.6cm}
\begin{tabular}{|c||c|c|c|} \hline
 $M^2_{U\alpha\beta}$
& \multicolumn{1}{c|}{\scriptsize{${\beta=1}$}} 
& \multicolumn{1}{c|}{\scriptsize{$\beta=2$}} 
& \multicolumn{1}{c|}{\scriptsize{$\beta=3$}} \\\hline\hline
 \scriptsize{$\alpha=1$}
& \multicolumn{1}{c|}{$(820)^2$} 
& \multicolumn{1}{c|}{0} 
& \multicolumn{1}{c|}{0} \\\hline

 \scriptsize{$\alpha=2$}
& \multicolumn{1}{c|}{0} 
& \multicolumn{1}{c|}{$(600)^2$} 
& \multicolumn{1}{c|}{$(373)^2$} \\\hline

 \scriptsize{$\alpha=3$}
& \multicolumn{1}{c|}{0} 
& \multicolumn{1}{c|}{$(373)^2$} 
& \multicolumn{1}{c|}{$(580)^2$} \\\hline
\end{tabular}
\vskip0.4cm
\caption{\label{tab1}
The basic MSSM parameters in our reference scenario with QFV.
All of $T_{U \a\b}$ and $T_{D \a\b}$ are set to zero.
All mass parameters are given in GeV. 
}
\end{center}
\end{table}

\begin{table}[t]
\begin{center}

\begin{tabular}{|c|c|c|c|c|c|} \hline
 
  \multicolumn{1}{|c|}{$\tilde u_1$} 
& \multicolumn{1}{c|}{$\tilde u_2$} 
& \multicolumn{1}{c|}{$\tilde u_3$} 
& \multicolumn{1}{c|}{$\tilde u_4$} 
& \multicolumn{1}{c|}{$\tilde u_5$} 
& \multicolumn{1}{c|}{$\tilde u_6$} \\\hline\hline
 
  \multicolumn{1}{|c|}{472} 
& \multicolumn{1}{c|}{708} 
& \multicolumn{1}{c|}{819} 
& \multicolumn{1}{c|}{837} 
& \multicolumn{1}{c|}{897} 
& \multicolumn{1}{c|}{918} \\\hline

\end{tabular}
\begin{tabular}{|c|c|c|c|c|c|} \hline
 
  \multicolumn{1}{|c|}{$\tilde d_1$} 
& \multicolumn{1}{c|}{$\tilde d_2$} 
& \multicolumn{1}{c|}{$\tilde d_3$} 
& \multicolumn{1}{c|}{$\tilde d_4$} 
& \multicolumn{1}{c|}{$\tilde d_5$} 
& \multicolumn{1}{c|}{$\tilde d_6$} \\\hline\hline
 
  \multicolumn{1}{|c|}{800} 
& \multicolumn{1}{c|}{820} 
& \multicolumn{1}{c|}{830} 
& \multicolumn{1}{c|}{835} 
& \multicolumn{1}{c|}{897} 
& \multicolumn{1}{c|}{922} \\\hline

\end{tabular}
\vskip0.2cm
%
\begin{tabular}{|c||c|c|c|c||c|c|} \hline
 
  \multicolumn{1}{|c||}{$\sg$} 
& \multicolumn{1}{c|}{$\tilde \chi^0_1$} 
& \multicolumn{1}{c|}{$\tilde \chi^0_2$} 
& \multicolumn{1}{c|}{$\tilde \chi^0_3$} 
& \multicolumn{1}{c||}{$\tilde \chi^0_4$} 
& \multicolumn{1}{c|}{$\tilde \chi^\pm_1$} 
& \multicolumn{1}{c|}{$\tilde \chi^\pm_2$} \\\hline\hline
 
  \multicolumn{1}{|c||}{800} 
& \multicolumn{1}{c|}{138} 
& \multicolumn{1}{c|}{261} 
& \multicolumn{1}{c|}{1003} 
& \multicolumn{1}{c||}{1007} 
& \multicolumn{1}{c|}{261} 
& \multicolumn{1}{c|}{1007} \\\hline

\end{tabular}
\begin{tabular}{|c|c|c|c|} \hline
 
  \multicolumn{1}{|c|}{$h^0$} 
& \multicolumn{1}{c|}{$H^0$} 
& \multicolumn{1}{c|}{$A^0$} 
& \multicolumn{1}{c|}{$H^\pm$} \\\hline\hline
 
  \multicolumn{1}{|c|}{122} 
& \multicolumn{1}{c|}{800} 
& \multicolumn{1}{c|}{800} 
& \multicolumn{1}{c|}{804} \\\hline

\end{tabular}
\vskip0.4cm
\caption{\label{tab2}
Sparticles, Higgs bosons and corresponding masses (in GeV) in the scenario of 
Table \ref{tab1}. $H^0$ is the heavier CP-even Higgs boson.
}
\end{center}
\end{table}

We take $\tan\b, m_{A^0}, M_1, M_2, M_3, \mu, M^2_{Q\a\b}, M^2_{U \a\b}, 
M^2_{D \a\b}, T_{U \a\b}$ and $T_{D \a\b}$ as the basic MSSM parameters at the 
weak scale and assume them to be real. 
Here $M_3$ is the SU(3) gaugino mass parameter. 
The QFV parameters are the squark generation 
mixing terms $M^2_{Q\a\b}$, $M^2_{U \a\b}$, $M^2_{D \a\b}$, $T_{U \a\b}$ and 
$T_{D \a\b}$ with $\a \neq \b$. 
We study a specific scenario which is chosen so that QFV signals at LHC may be maximized and 
hence can serve as a benchmark scenario for further experimental investigations.
As such a scenario, we take the scenario specified by Table \ref{tab1}, which was 
studied for QFV gluino decays in \cite{Bartl:QFV_gluino_decay}. 
Here we take $M_1=(5/3)\tan^2\t_W M_2$, assuming gaugino mass unification including the 
gluino mass parameter $M_3$.
%
In this scenario one has $\delta^{uLL}_{23}=0.068$, $\delta^{uRR}_{23}=0.4$ and 
$\delta^{uRL}_{23}=\delta^{uRL}_{32}=0$ for the QFV parameters. 
%
This scenario satisfies the conditions (i)-(v). 
For the observables in (i) and (ii) we obtain $B(b \to s \gamma)=3.56\times10^{-4}, 
~B(b \to s l^+l^-)= 1.59\times10^{-6}, 
~B(B_s \to \mu^+\mu^-)=4.71\times10^{-9}, ~B(B_u^- \to \tau^- {\bar\nu}_\tau)=7.85\times10^{-5}, 
~\Delta M_{B_s}= 17.37 ~ps^{-1}$ and $\Delta\rho(SUSY)= 1.51\times10^{-4}$. 
The resulting tree-level masses of squarks, neutralinos and charginos are given in Table 
\ref{tab2} and the up-type squark compositions in the flavour eigenstates in 
Table \ref{tab3}.
\begin{table}[t]
\begin{center}
\begin{tabular}{|c||c|c|c|c|c|c|}
		 \hline
			$|R^{\tilde u}_{i\a}|$
		  & $\ti u_L$ & $\ti c_L$ & $\ti t_L$ & $\ti u_R$ & $\ti c_R$ & $\ti t_R$ \\
		 \hline\hline
		  $\ti u_1$  & 0.001 & 0.004 & 0.024 & 0 &  0.715 & 0.699 \\
		  $\ti u_2$  & 0.003 & 0.014 & 0.055 & 0 & 0.699 & 0.713 \\
		  $\ti u_3$  & 0 & 0 & 0 & 1.0 & 0 & 0 \\
		  $\ti u_4$  & 0.128 & 0.584 & 0.800 & 0 & 0.021 & 0.053 \\
		  $\ti u_5$  & 0.181 & 0.781 & 0.598 & 0 & 0.008 & 0.024 \\
		  $\ti u_6$  & 0.975 & 0.221 & 0.005 & 0 & 0 & 0 \\
%
%

		 \hline
		 \end{tabular} 
		 
\vspace{3mm} 
%
%

	\caption{\label{tab3} 
The up-type squark compositions in the flavour eigenstates, 
i.e. the absolute values of the mixing matrix elements $R^{\ti u}_{i\a}$ 
for the scenario of Table \ref{tab1}.}
%

\end{center}
\end{table}

For the most important decay branching ratios of the two lightest up-type squarks we get 
$B(\ti{u}_1 \to c \nt_1)= 0.59, ~B(\ti{u}_1 \to t \nt_1)= 0.39, 
~B(\ti{u}_2 \to c \nt_1)= 0.44, ~B(\ti{u}_2 \to t \nt_1)= 0.40$. 
Note that the branching ratios of the decays of a squark into quarks of different 
generations are very large simultaneously, which could lead to large QFV effects. 
In our scenario this is a consequence of the facts that both squarks $\su_{1,2}$ are 
mainly strong mixtures of $\tilde c_R$ and $\tilde t_R$ due to the large 
$\tilde c_R -\tilde t_R$ mixing term $M^2_{U 23} (= (373 \gev)^2)$ (see Table 3) 
and that $\nt_1$ is mainly the $U(1)$ gaugino. 
This also suppresses the couplings of $\su_{1,2}$ to $\tilde \chi^0_2$ and 
$\tilde \chi^+_1$ which are mainly $SU(2)$ gauginos. 
Note that $\nt_{3,4}$ and $\ch_2$ are very heavy in this scenario.\\
%
The main decay branching ratios of the other up-type squarks are as follows:
$B(\ti{u}_3 \to u \nt_1)=0.93, ~B(\ti{u}_4 \to c \nt_2)=0.09, ~B(\ti{u}_4 \to t \nt_2)=0.21, 
~B(\ti{u}_4 \to s \chp_1)=0.21, ~B(\ti{u}_4 \to b \chp_1)=0.45, ~B(\ti{u}_5 \to c \nt_2)=0.19, 
~B(\ti{u}_5 \to t \nt_2)=0.07, ~B(\ti{u}_5 \to s \chp_1)=0.37, ~B(\ti{u}_5 \to b \chp_1)=0.17, 
~B(\ti{u}_5 \to c \sg)=0.17, ~B(\ti{u}_6 \to u \nt_2)=0.22, ~B(\ti{u}_6 \to d \chp_1)=0.47$, and 
$~B(\ti{u}_6 \to u \sg)=0.28$.

We now study various parameter dependences of the QFV squark decay 
branching ratios for the reference scenario of Table \ref{tab1}. 
In all plots we mark the point corresponding to this scenario by an "x". 
In Figs.1-3 we show that both $B(\su_i \to c \nt_1)$ and 
$B(\su_i \to t \nt_1)$ (i=1,2) can be very large simultaneously in a sizable 
QFV parameter region satisfying all of the conditions (i)-(v), which can lead 
to large rates for QFV signal events at LHC as we will see in the next section.
%
%
%
%

Fig.1 shows the contours of $B(\su_1 \to c \nt_1)$ and $B(\su_1 \to t \nt_1)$ 
in the ($\Delta M^2_U, M^2_{U 23}$) plane with $\Delta M^2_U \equiv 
M^2_{U 22} - M^2_{U 33}$. The range of $M^2_{U23}$ shown corresponds to the 
range $|\delta^{uRR}_{23}| < 0.45$ for $\Delta M^2_U=0$. 
In the region shown all of the low energy constraints are fulfilled. 
We see that there are sizable regions where both decay modes are important 
at the same time. The observed behaviour can be easily understood in the 
limit where the $\st_L$ - $\st_R$ mixing is neglected since in this limit only 
the mixing between $\sc_R$ and $\st_R$ is relevant for $\su_{1,2}$ and the 
corresponding effective mixing angle is given by  
$\tan(2\theta^{eff}_{\sc_R \st_R}) \equiv 2M^2_{U 23}/(\Delta M^2_U - m^2_t)$.
Note that for $\Delta M^2_U - m^2_t > 0$ [$\Delta M^2_U  - m^2_t < 0$], we have  
$\su_1$ $\sim$ $\st_R$ (+ $\sc_R$) [$\su_1$ $\sim$ $\sc_R$ (+ $\st_R$)].
We also find that the behaviour of $B(\su_2 \to c \nt_1)$ and $B(\su_2 \to t \nt_1)$ 
is similar to that of $B(\su_1 \to t \nt_1)$ and $B(\su_1 \to c \nt_1)$, respectively, 
which is a consequence of the fact that mainly the mixing between $\sc_R$ and $\st_R$ 
is important for the $\su_{1,2}$ system.
\begin{figure}
\begin{center}
\scalebox{0.5}[0.8]{\includegraphics{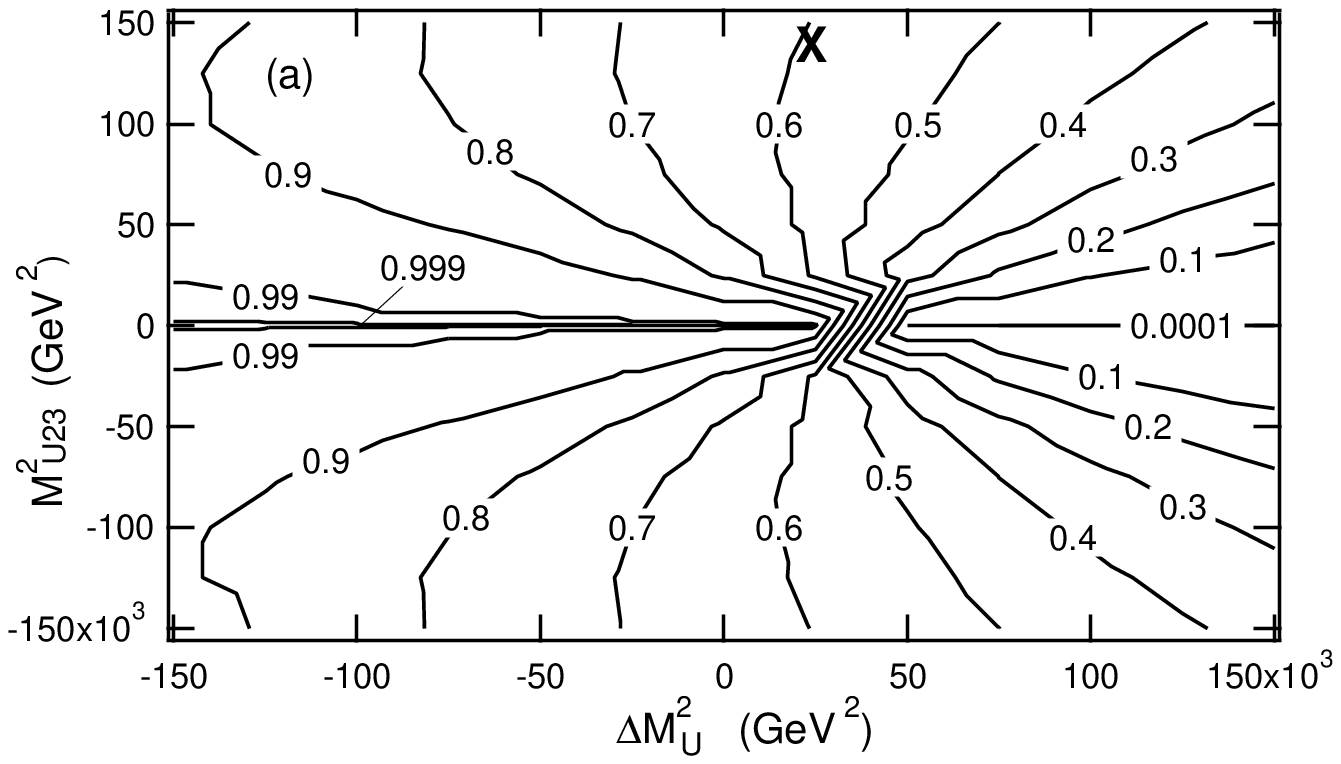}}
\scalebox{0.5}[0.8]{\includegraphics{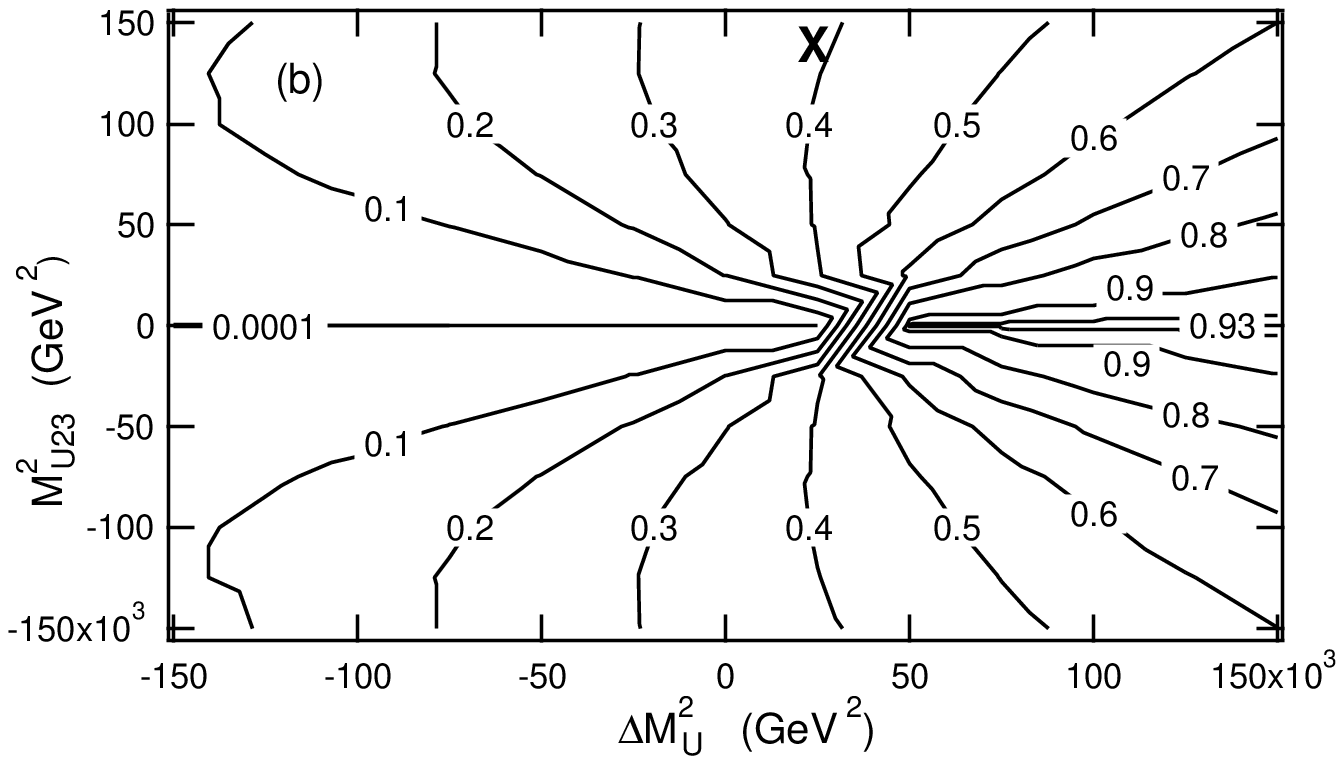}}
\caption{
Contours of the QFV decay branching ratios (a)$B(\su_1 \to c \nt_1)$ and 
(b)$B(\su_1 \to t \nt_1)$ in the ($\Delta M^2_U, M^2_{U 23}$) plane where 
all of the conditions (i)-(v) are satisfied. 
}
\label{fig1}
\end{center}
\end{figure}

Fig.2 presents contours of $B(\su_2 \to c \nt_1)$ and 
$B(\su_2 \to t \nt_1)$ in the $\delta^{uLL}_{23} - \delta^{uRR}_{23}$ 
plane where all of the conditions (i)-(v) are satisfied except the $b \to s \gamma$ 
constraint which we show by plotting the corresponding $B(b \to s \gamma)$ contours. 
All basic parameters other than $M^2_{Q 23}$ and $M^2_{U 23}$ are fixed as in the 
scenario of Table \ref{tab1}. 
For $B(\su_1 \to c \nt_1)$ and $B(\su_1 \to t \nt_1)$ we have obtained similar 
contours to Fig.2.(b) and Fig.2.(a), respectively, but they are almost flat.
From Fig.2 we find that the possibility of the large 
QFV effect cannot be excluded by the $b \to s \gamma$ constraint even if the 
experimental error of $B(b \to s \gamma)$ becomes very small.
We see also that $B(\su_2 \to c \nt_1)$ and $B(\su_2\to t \nt_1)$ are 
sensitive [rather insensitive] to $\delta^{uRR}_{23}$ [$\delta^{uLL}_{23}$].
For large values of $\delta^{uRR}_{23}$ we see that there is a mild 
dependence on $\delta^{uLL}_{23}$. This is due the fact that 
for large $\delta^{uRR}_{23}$ the mass squared difference between $\su_2$ 
(the heavier of the RR sector, i.e. the $\sc_R$-$\st_R$ sector) and 
$\su_4$ (the lighter of the LL sector, i.e. the $\sc_L$-$\st_L$ sector) 
becomes small and of the same size as the $\st_L$-$\st_R$ mixing term 
(= $-m_t \mu \cot\beta$ (see Eq.(\ref{eq:RL}))) enhancing the mixing between 
the RR and LL sectors. 
For small values of $\delta^{uRR}_{23}$ the RR sector decouples effectively 
from the LL sector and hence the $\su_2$ decay branching ratios are almost 
independent of $\delta^{uLL}_{23}$. 
\begin{figure}
\begin{center}
\scalebox{0.48}[0.6]{\includegraphics{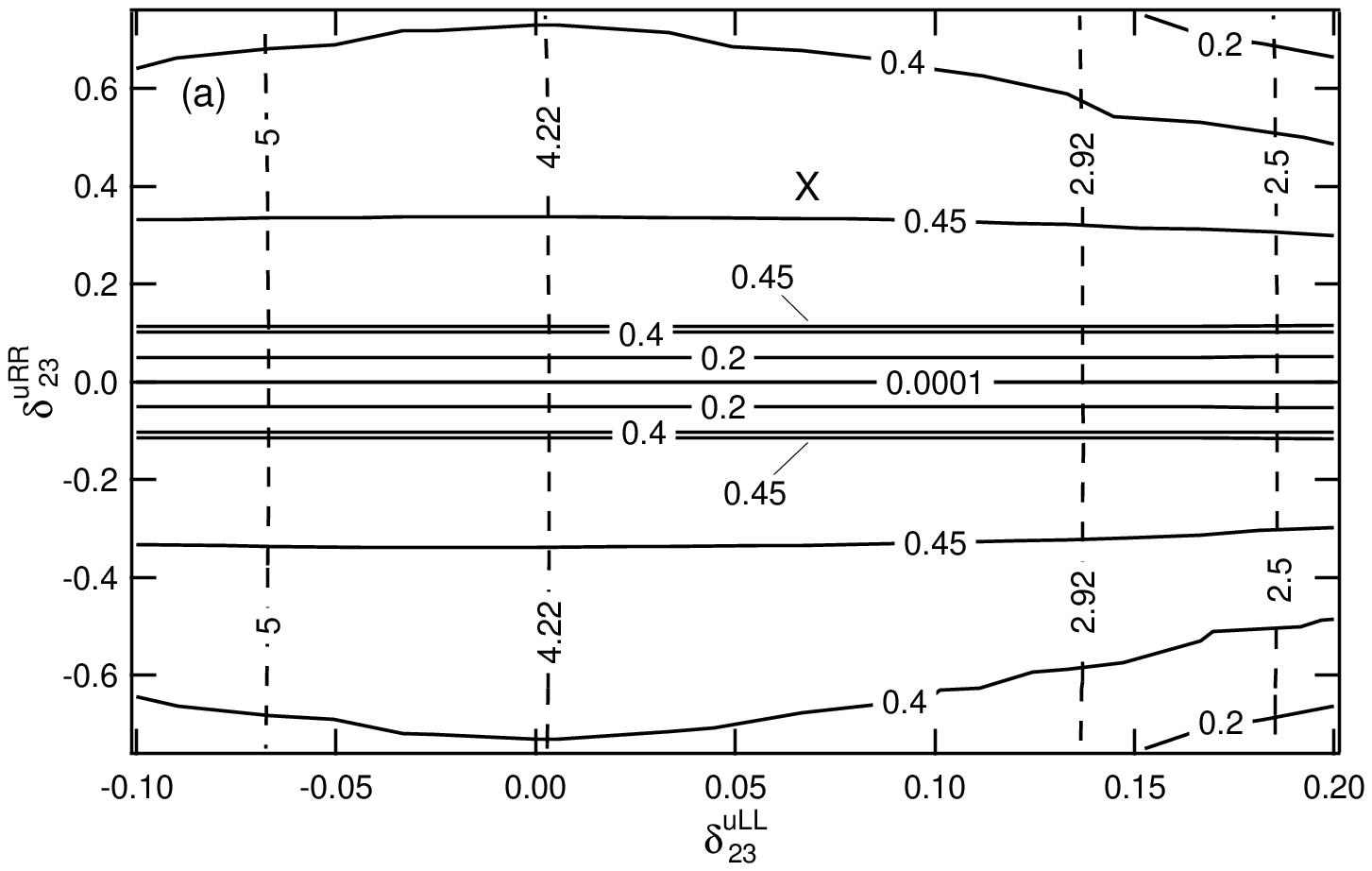}}
\scalebox{0.48}[0.6]{\includegraphics{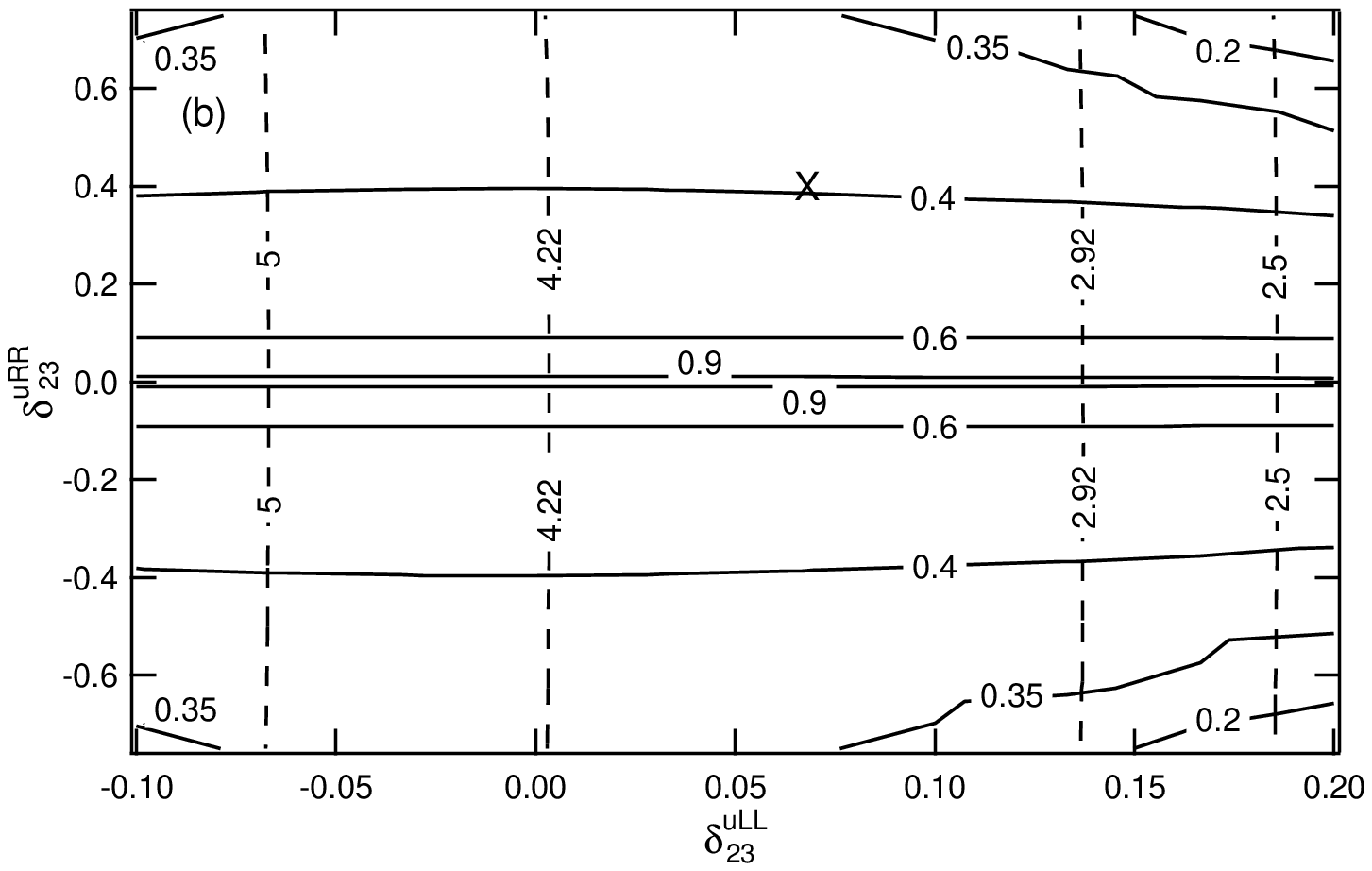}}\\
%
\caption{
Contours of (a)$B(\su_2 \to c \nt_1)$ and (b)$B(\su_2 \to t \nt_1)$ (solid lines) 
in the $\delta^{uLL}_{23} - \delta^{uRR}_{23}$ plane where all of the conditions (i)-(v) 
except the $b \to s \gamma$ constraint are satisfied. 
Contours of $10^4 \times B(b \to s \gamma)$ (dashed lines) are also shown. 
The condition (i) requires $2.92 < 10^4 \times B(b \to s ~\gamma) < 4.22$. 
}
\label{fig2}
\end{center}
\end{figure}

In Fig.3 we show the $\delta^{uRL}_{23}$ dependences of the $\su_{1,2}$ decay 
branching ratios, where all basic parameters other than $T_{U 32}$ are fixed 
as in the scenario of Table \ref{tab1}. 
The observed dependences are a consequence of the enhanced $\st_L$ component 
in $\su_{1,2} (\sim {\ti c}_R + {\ti t}_R)$ for increased $|\delta^{uRL}_{23}|$. 
The enhanced $\st_L$ content implies an enhancement of the $b ~\chp_1(\simeq {\ti W}^+)$ mode. 
The enhancement of $B(\su_2 \to \su_1 ~h^0)$ for increased $|\delta^{uRL}_{23}|$ 
is partly also caused by the enhanced $\st_L$ component and, more importantly, 
by the increased coupling of $\su_2 \su_1 h^0$ which contains a term proportional 
to $T_{U32}$.  Note that in such scenarios squark decays could be additional 
sources of the Higgs boson. 
The asymmetry with respect to $\delta^{uRL}_{23}=0$ follows from 
the $\st_L$ - $\st_R$ mixing term (= $-m_t \mu \cot\beta \ne 0$ (see Eq.(\ref{eq:RL}))) 
which already induces some $\st_L$ component in $\su_{1,2}$ (see Table \ref{tab3}). 
\begin{figure}
\begin{center}
\scalebox{0.6}[0.6]{\includegraphics{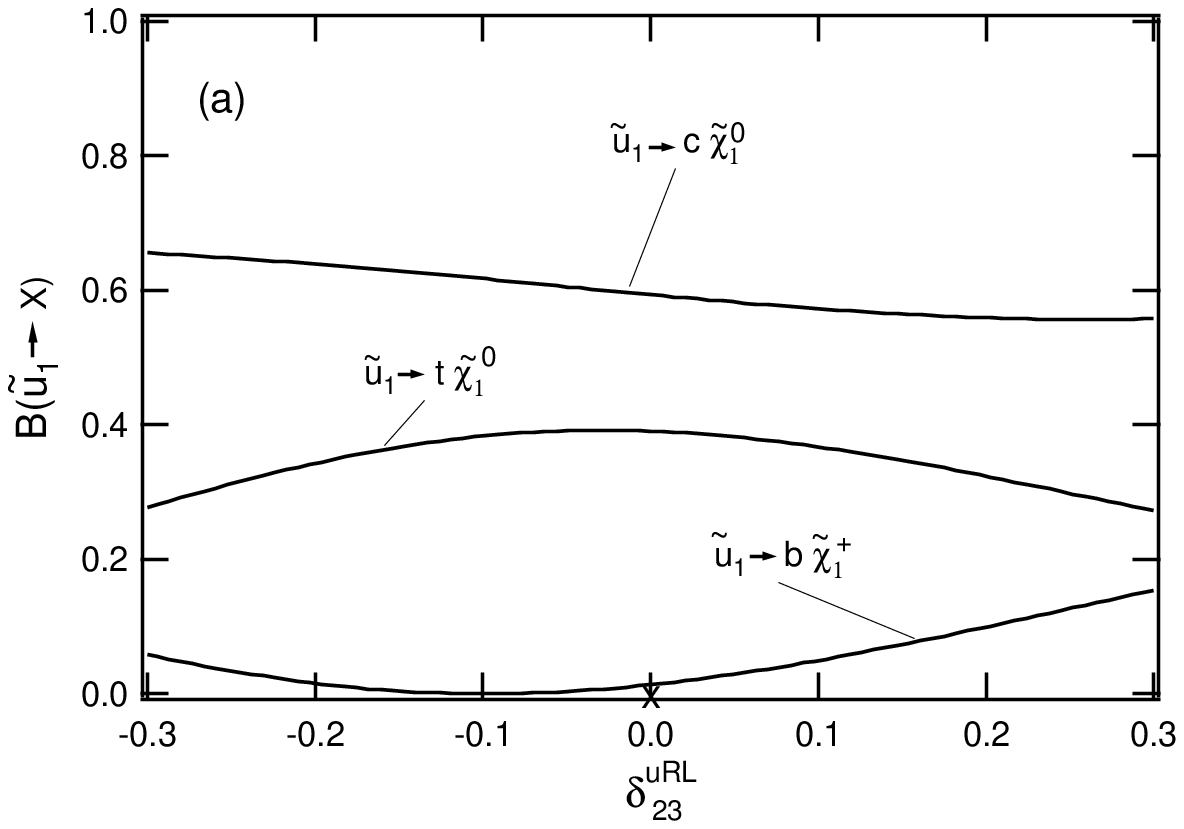}}
\scalebox{0.6}[0.6]{\includegraphics{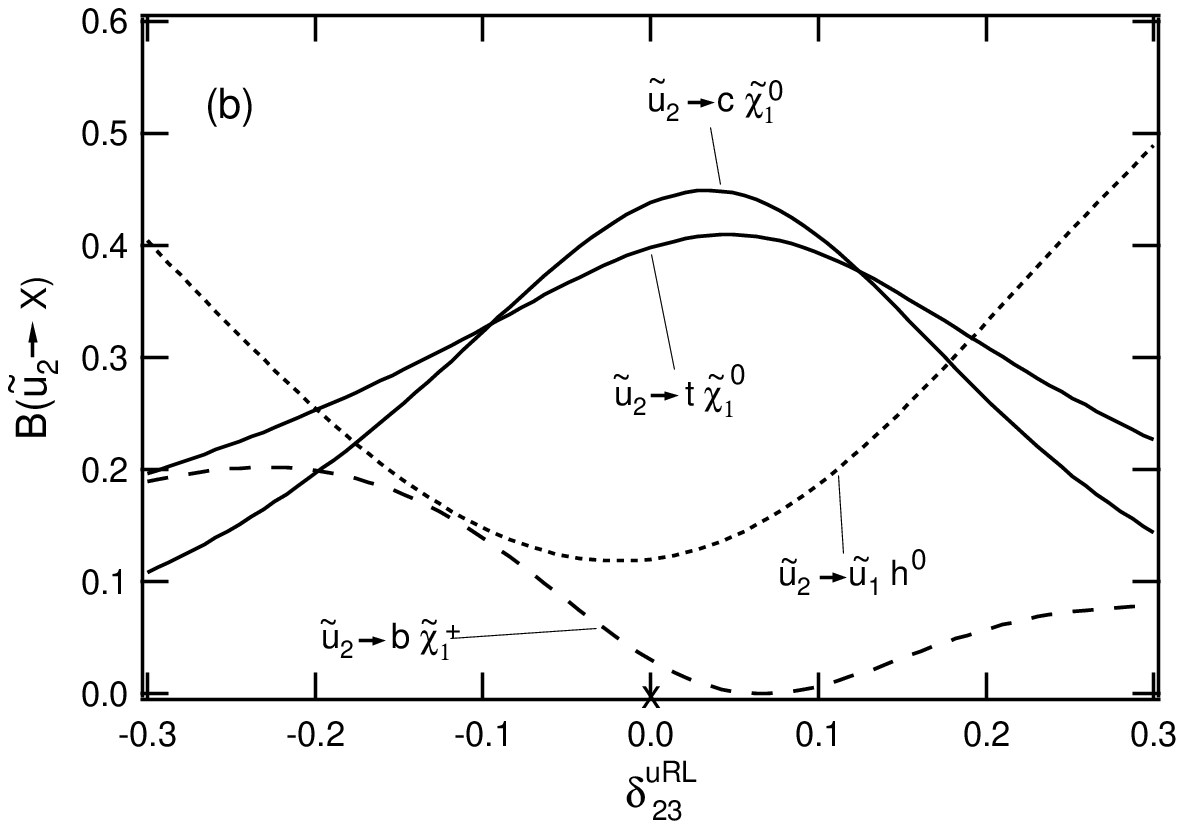}}
\caption{
$\delta^{uRL}_{23}$ dependences of the (a) $\su_1$ and (b) $\su_2$ decay branching ratios. 
The shown range of $\delta^{uRL}_{23}$ 
is the whole range allowed by the conditions (i) to (v) given in the text; note that the 
range $|\delta^{uRL}_{23}| \gsim 0.3$ is excluded by the condition (v).
}
\label{fig3}
\end{center}
\end{figure}
%
As for the $\delta^{uRL}_{32}$ 
dependence of the $\su_{1,2}$ decay branching ratios, 
we have obtained results similar to those 
for the $\delta^{uRL}_{23}$ dependence in Fig.3. 
\section{Impact on collider signatures}
%
%

%
%
We now study effects of the squark generation mixing on QFV signals at LHC. 
The large $B(\su_i \to c \nt_1)$ and $B(\su_i \to t \nt_1)$ $(i=1,2)$ may result 
in a sizable rate for the following QFV signals: 
\beq \label{eq:QFVsq_production_decay}
p~p \to \su_{1,2} ~\bar{\su}_{1,2} ~X \to c ~\bar{t} ~\nt_1 ~\nt_1 ~X, ~t ~\bar{c} ~\nt_1 ~\nt_1 ~X,
\eeq
where X contains only beam-jets and the $\nt_1$'s give rise to missing transverse energy $E_T^{mis}$.
The corresponding cross sections are given by 
\begin{eqnarray}
   \hspace{-1.0cm} 
   \sigma^{ij}_{ct} & \hspace{-0.2cm} \equiv & \hspace{-0.2cm} 
   \sigma(pp \to \su_i \bar{\su}_j X \to c \bar{t} (t \bar{c}) \nt_1 \nt_1 X)
   \nonumber\\ 
   \hspace{-1.0cm}
               & \hspace{-0.2cm} \equiv & \hspace{-0.2cm}  
   \sigma(pp \to \su_i \bar{\su}_j X \to c \bar{t} \nt_1 \nt_1 X)
   + \sigma(pp \to \su_i \bar{\su}_j X \to t \bar{c} \nt_1 \nt_1 X)
   \nonumber\\ 
   \hspace{-1.0cm}
               & \hspace{-0.2cm}      = & \hspace{-0.2cm} 
   \sigma(pp \to \su_i \bar{\su}_j X) [B(\su_i \to c \nt_1) \cdot B(\bar{\su}_j \to \bar{t} \nt_1) 
                                     + B(\su_i \to t \nt_1) \cdot B(\bar{\su}_j \to \bar{c} \nt_1)].
   \label{eq: ct production cross section} 
\end{eqnarray}
%
%
%
We calculate the relevant squark-squark and squark-antisquark pair  
production cross sections at leading order using the WHIZARD/O'MEGA packages \cite{Whizard, Omega} 
where we have implemented the model described in Section 2 with squark 
generation mixing in its most general form. 
We use the CTEQ6L global parton density fit \cite{CTEQ6} for the parton distribution 
functions and take $Q = m_{\su_i} + m_{\su_j}$ for the factorization scale, where 
$\su_i$ and $\su_j$ are the squark pair produced.
The QCD coupling $\alpha_s(Q)$ is also evaluated (at the two-loop level) at 
this scale Q. We have cross-checked our implementation of QFV by comparing with 
the results obtained using the public packages FeynArts \cite{FeynArts} and 
FormCalc \cite{FormCalc}.

Defining QFC production cross sections as 
\begin{eqnarray}
  \sigma^{ij}_{q\bar{q}} & \hspace{-0.2cm} \equiv & \hspace{-0.2cm}
  \sigma(pp \to \su_i \bar{\su}_j X \to q \bar{q} \nt_1 \nt_1 X) 
  \nonumber\\ 
                         & \hspace{-0.2cm} = & \hspace{-0.2cm}  
  \sigma(pp \to \su_i \bar{\su}_j X) \cdot B(\su_i \to q \nt_1) 
                           \cdot B(\bar{\su}_j \to \bar{q} \nt_1) ~ \quad (q=c,t),
  \label{eq:QFC_squark_production_decay}
\end{eqnarray}
we obtain the following cross sections 
at the center-of-mass energy $E_{cm}$=14 TeV [7 TeV] in the scenario of Table \ref{tab1}: 
$\sigma^{11}_{ct}$ = 172.8 [11.8] fb, $\sigma^{22}_{ct}$ = 11.5 [0.41] fb, 
$\sigma^{11}_{c\bar{c}}$ = 131.4 [9.0] fb, $\sigma^{22}_{c\bar{c}}$ = 6.3 [0.23] fb, 
$\sigma^{11}_{t\bar{t}}$ = 56.8 [3.89] fb, $\sigma^{22}_{t\bar{t}}$ = 5.2 [0.19] fb. 
%
%
The expected number of the $c \bar{t} \, / \, t \bar{c}$ production events of 
Eq. (\ref{eq:QFVsq_production_decay}) is ${\mathcal L} \cdot \sum_{i,j=1,2} \sigma^{ij}_{ct} \simeq$ 18400
 [10] events for an integrated luminosity of ${\mathcal L}=100 fb^{-1} [1 fb^{-1}]$ at LHC 
with $E_{cm}$=14 TeV [7 TeV].\\
%
The main contribution to $\sigma(pp \to \su_i \bar{\su}_i X)$ $(i=1,2)$ comes from the 
subprocess $g g \to \su_i \bar{\su}_i$. The gluon-$\su_i$-$\su_j$ coupling vanishes 
for $i \neq j$ due to the color SU(3) symmetry. 
Therefore, $\sigma(pp \to \su_i \bar{\su}_j X)$ and hence $\sigma^{ij}_{ct}$, 
$\sigma^{ij}_{c\bar{c}}$ and $\sigma^{ij}_{t\bar{t}}$ are very small for $i \neq j$, 
e.g. O(0.01) fb [O($10^{-4}$) fb] for $(i,j)=(1,2)$ at $E_{cm}$ =14 TeV [7 TeV].
We have found that the production cross sections of the quark 
pair ($c \bar t$, $t\bar c$, $c\bar c$, $t\bar t$) plus two $\nt_1$'s and $n ~\nu$'s 
($n=0,2,4,\dots$) via production of the heavier up-type squarks $\su_i$ 
($i \geq 3$) are very small in this scenario. 

In Fig.4 we show the $\delta^{uRR}_{23}$ 
dependences of the QFV production cross sections $\sigma^{ii}_{ct}$ $(i=1,2)$ at $E_{cm}$ = 7 
and 14 TeV, where all basic parameters other than $M^2_{U 23}$ are fixed as in the scenario of 
Table \ref{tab1}. The QFV cross sections at 14 TeV are about an order of magnitude larger 
than those at 7 TeV. 
We see that the QFV cross sections quickly increase with increase of 
the QFV parameter $|\delta^{uRR}_{23}|$ around $\delta^{uRR}_{23} = 0$ and that they can be 
quite sizable in a wide allowed range of $\delta^{uRR}_{23}$. The mass of $\su_1$ ($\su_2$) 
decreases (increases) with increase of $|\delta^{uRR}_{23}|$. This leads to the increase of 
$\sigma^{11}_{ct}$ and the decrease of $\sigma^{22}_{ct}$ with increase of $|\delta^{uRR}_{23}|$.
$\sigma^{11}_{ct}$ vanishes for $|\delta^{uRR}_{23}| \gsim 0.76$, where the decay 
$\su_1 \to t \nt_1$ is kinematically forbidden. 
We have $\su_2 = \su_R$ for $|\delta^{uRR}_{23}| 
\gsim 0.9$, which explains the enhancement of $\sigma(pp \to \su_2 \bar{\su}_2 X)$  and the 
vanishing of $\sigma^{22}_{ct}$ for $|\delta^{uRR}_{23}| \gsim 0.9$. Note that in case 
$\su_2 = \su_R$, the subprocess $u \bar{u} \to \su_2(=\su_R) \bar{\su}_2(=\bar{\su}_R)$ via 
t-channel gluino exchange also can contribute to $\sigma(pp \to \su_2 \bar{\su}_2 X)$. 
\begin{figure}
\begin{center}
\scalebox{0.5}[0.6]{\includegraphics{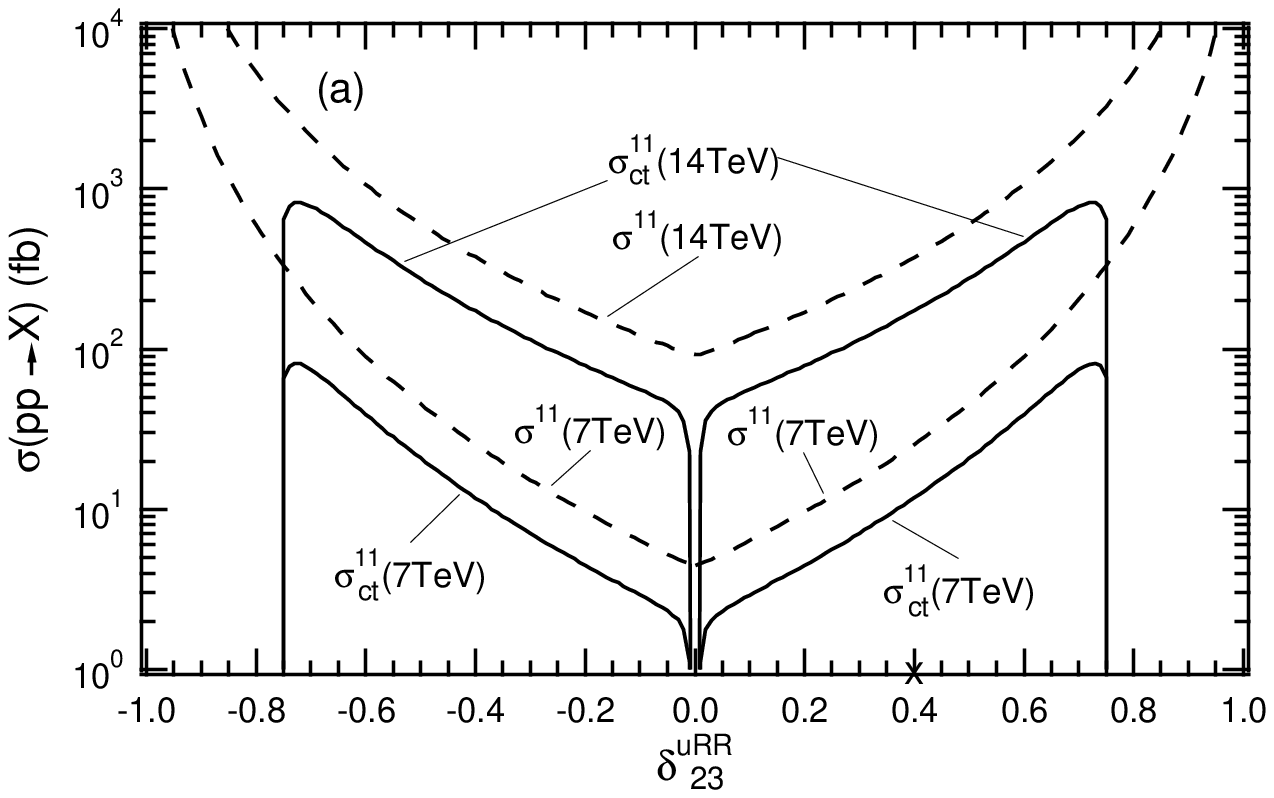}}
\scalebox{0.47}[0.6]{\includegraphics{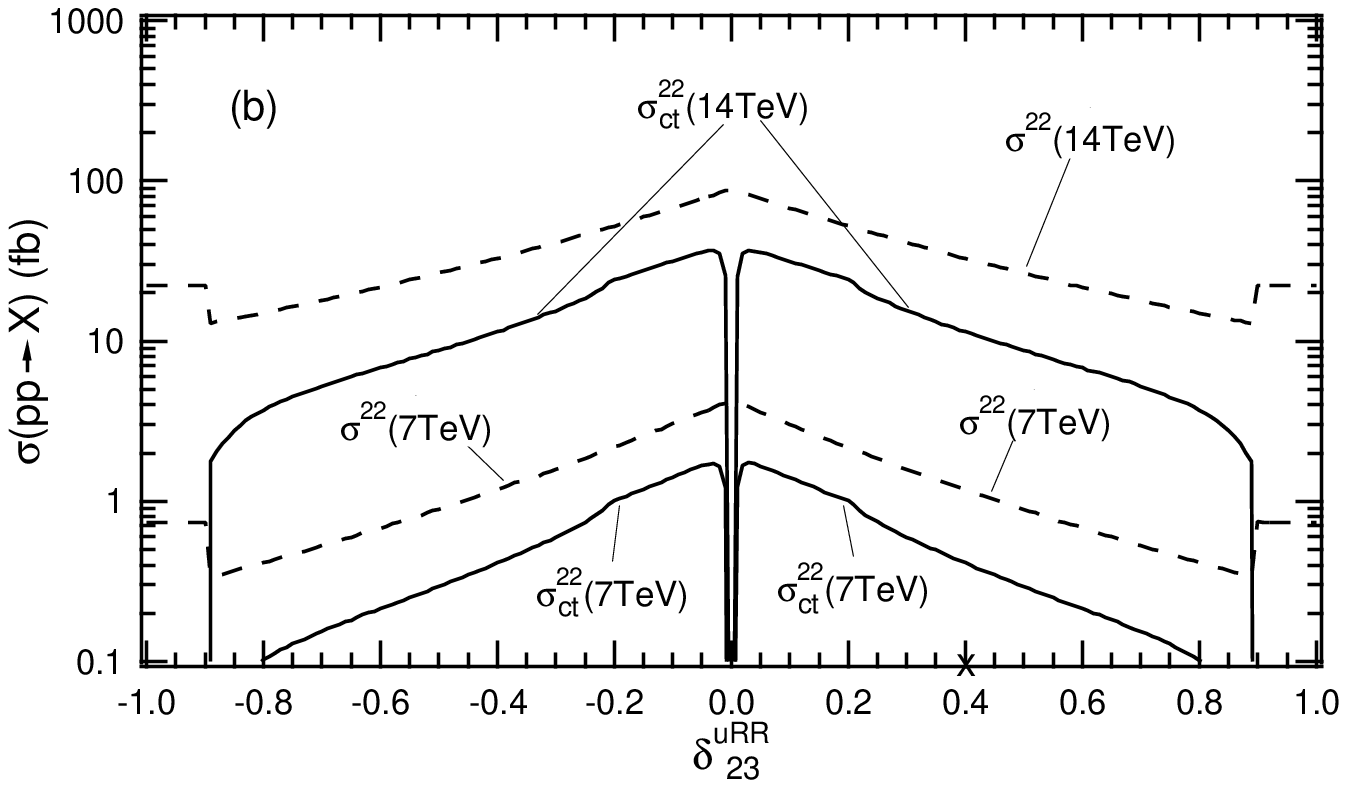}}
\caption{
$\delta^{uRR}_{23}$ dependences of (a) $\sigma^{11} \equiv \sigma(pp \to \su_1 \bar{\su}_1 X)$, 
$\sigma^{11}_{ct}$ and (b) $\sigma^{22} \equiv \sigma(pp \to \su_2 \bar{\su}_2 X)$, 
$\sigma^{22}_{ct}$ at $E_{cm}$ = 7 and 14 TeV. 
The point "x" of $\delta^{uRR}_{23} = 0.4$ 
corresponds to our reference scenario of Table \ref{tab1}. 
The shown range of $\delta^{uRR}_{23}$ is the whole 
range allowed by the conditions (i) to (v) given in the text; note that the range 
$|\delta^{uRR}_{23}| \gsim 1.0$ is excluded by the condition $m_{\ti{u}_1} > m_{\nt_1}$ 
in (iii).
}
\label{fig4}
\end{center}
\end{figure}
%

We have also studied the $\delta^{uRL}_{32}$ dependence of the 
QFV production cross sections $\sigma^{ii}_{ct}$ $(i=1,2)$ at $E_{cm}$ = 7 TeV and 14 TeV, 
where all basic parameters other than $T_{U23}$ are fixed as in the scenario of 
Table \ref{tab1}. We find that the QFV cross sections are rather insensitive to the QFV 
parameter $\delta^{uRL}_{32}$ and that they can be large in a 
wide allowed range $|\delta^{uRL}_{32}| \lsim 0.3$ : 
$\sigma^{11}_{ct} \sim $ 170 [10] fb, $\sigma^{22}_{ct} \sim $ 10 [0.4] fb 
at $E_{cm}$=14 TeV [7 TeV]. 
The masses of $\su_{1,2}$ decrease (and hence the cross sections $\sigma^{ii}$ 
$(i=1,2)$ increase) and the branching ratios $B(\su_{1,2} \to c/t ~\nt_1)$ tend to 
decrease with increase of $|\delta^{uRL}_{32}|$. This implies that the QFV cross sections 
are rather insensitive to $\delta^{uRL}_{32}$. 
As for the $\delta^{uRL}_{23}$ dependence of $\sigma^{ii}_{ct}$ $(i=1,2)$ we have 
obtained similar results to those for the $\delta^{uRL}_{32}$ dependence.
%
%


The large $\sc_R$ - $\st_R$ mixing could also give rise to the following QFV production 
cross sections: 
%
\vspace{-0.5cm}
\begin{eqnarray}
  \sigma^{ij}_{tt} & \hspace{-0.2cm} \equiv & \hspace{-0.2cm}
  \sigma(pp \to \su_i ~\su_j ~X \to t ~t ~\nt_1 ~\nt_1 ~X) 
  \nonumber\\ 
%
                         & \hspace{-0.2cm} = & \hspace{-0.2cm}  
  \sigma(pp \to \su_i ~\su_j ~X) \cdot B(\su_i \to t ~\nt_1) 
                           \cdot B(\su_j \to t ~\nt_1) ~ (i,j=1,2)~, 
\end{eqnarray}
where X contains only beam-jets.  Here the $\su_i ~\su_j$ pair 
(with $\su_{i,j} \sim \sc_R + \st_R$ in the scenario under consideration) is 
produced mainly via a t-channel gluino exchange subprocess $ c ~c \to \su_i ~\su_j$ 
with c being the charm-quark in the beam proton. Note that the signal event 
"top-quark + top-quark + $E_T^{mis}$ + beam-jets" can practically not be produced in 
the MSSM with QFC (nor in the SM). 
It turns out however that in the scenario of Table \ref{tab1} the corresponding 
cross section $\sigma_{tt} \equiv \sigma^{11}_{tt}+\sigma^{12}_{tt}$ is at most 
O(0.1) fb at $E_{cm}$=14 TeV and hence that it might be relevant for a very high 
luminosity \cite{high luminosity LHC}. Therefore this QFV process will not be 
discussed further. 

In addition, we study QFV in production and decays of squarks at LHC 
for a QFV scenario based on the mSUGRA scenario SPS1a' \cite{SPS1a'} which 
has served as input for several experimental studies. 
The high energy inputs at the GUT scale $M_{GUT}=2.47 \times 10^{16}$ GeV in the scenario SPS1a' 
are taken as $m_0$=70 GeV, $m_{1/2}$=250 GeV, $A_0=-300$ GeV and $\mu > 0$ together with 
$\tan\beta(m_Z)$=10. Here $m_0$, $m_{1/2}$ and $A_0$ are the common scalar mass, gaugino mass 
and trilinear coupling at the GUT scale, respectively.
We use SPheno v3.0 \cite{SPheno_B-physics_refs} to obtain the resulting MSSM 
parameters at the scale Q=1 TeV according to the SPA convention \cite{SPS1a'}. 
At this scale, we add the QFV parameters (i.e. the squark generation mixing 
parameters) and vary them around zero (i.e. around the MFV scenario). 
An example set of the MSSM parameters thus obtained is given in Table 4 and 
the resulting mass spectrum and the up-type squark compositions in the flavour 
eigenstates in Tables 5 and 6, respectively. 
In this scenario one has $\delta^{uLL}_{23}=0$, $\delta^{uRR}_{23}=0.4$ 
and $\delta^{uRL}_{23}=\delta^{uRL}_{32}=0$ for the QFV parameters at the scale Q=1 TeV. 
Note that the resulting squark and gluino masses are smaller than 
those in the scenario of Table 1. We have checked that all of the constraints in 
Section 3 are fulfilled in this scenario.
For the important squark decay branching ratios we obtain 
$B(\su_1 \to c \nt_1)= 0.100, ~B(\su_1 \to t \nt_1)= 0.230, 
B(\su_2 \to c \nt_1)= 0.146, ~B(\su_2 \to t \nt_1)= 0.004$. 
In this scenario the squark mass eigenstate $\su_1$ ($\su_2$) 
is dominated by a strong mixture of the flavour eigenstates $\st_R$, $\st_L$ 
and $\sc_R$ ($\st_L$ and $\sc_R$) and $\nt_1$ is nearly the U(1) gaugino $\ti B^0$ 
which couples to the right up-type squarks sizably. 
This explains the sizable branching ratios of $B(\su_1 \to c \nt_1)$, 
$B(\su_1 \to t \nt_1)$ and $B(\su_2 \to c \nt_1)$ and the very small 
$B(\su_2 \to t \nt_1)$. 
In this scenario we obtain the following cross sections at the center-of-mass energy 
$E_{cm}$=14 TeV [7 TeV]: $\sigma^{11}_{ct}$ = 119.7 [11.8] fb, 
$\sigma^{22}_{ct}$ = 0.197 [0.01] fb. 
Note that the QFV decay branching ratios $B(\su_{1,2} \to c/t ~\nt_1)$ are 
significantly smaller than those in the scenario of Table 1, but that the QFV 
production cross section $\sigma^{11}_{ct}$ is nevertheless large due to the 
lighter squarks in this scenario based on SPS1a'. 

\begin{table}[t]
\begin{center}

\begin{tabular}{|c||c|c|c|} \hline
 $M^2_{Q\alpha\beta}$
& \multicolumn{1}{c|}{\scriptsize{${\beta=1}$}} 
& \multicolumn{1}{c|}{\scriptsize{$\beta=2$}} 
& \multicolumn{1}{c|}{\scriptsize{$\beta=3$}} \\\hline\hline
 \scriptsize{$\alpha=1$}
& \multicolumn{1}{c|}{$(526)^2$} 
& \multicolumn{1}{c|}{0} 
& \multicolumn{1}{c|}{0} \\\hline

 \scriptsize{$\alpha=2$}
& \multicolumn{1}{c|}{0} 
& \multicolumn{1}{c|}{$(526)^2$} 
& \multicolumn{1}{c|}{0} \\\hline

 \scriptsize{$\alpha=3$}
& \multicolumn{1}{c|}{0} 
& \multicolumn{1}{c|}{0} 
& \multicolumn{1}{c|}{$(471)^2$} \\\hline
\end{tabular}
\hspace{0.1cm}
\begin{tabular}{|c|c|c|c|c|c|} \hline
 
  \multicolumn{1}{|c|}{$M_1$} 
& \multicolumn{1}{c|}{$M_2$} 
& \multicolumn{1}{c|}{$M_3$} 
& \multicolumn{1}{c|}{$\mu$} 
& \multicolumn{1}{c|}{$\tan\beta$} 
& \multicolumn{1}{c|}{$m_{A^0}$} \\\hline\hline
 
  \multicolumn{1}{|c|}{103} 
& \multicolumn{1}{c|}{193} 
& \multicolumn{1}{c|}{572} 
& \multicolumn{1}{c|}{398} 
& \multicolumn{1}{c|}{10} 
& \multicolumn{1}{c|}{373} \\\hline\hline
 
  \multicolumn{1}{|c|}{$T_{U11}$} 
& \multicolumn{1}{c|}{$T_{U22}$} 
& \multicolumn{1}{c|}{$T_{U33}$} 
& \multicolumn{1}{c|}{$T_{D11}$} 
& \multicolumn{1}{c|}{$T_{D22}$} 
& \multicolumn{1}{c|}{$T_{D33}$} \\\hline\hline
 
  \multicolumn{1}{|c|}{-0.007} 
& \multicolumn{1}{c|}{-2.68} 
& \multicolumn{1}{c|}{-488} 
& \multicolumn{1}{c|}{-0.19} 
& \multicolumn{1}{c|}{-3.26} 
& \multicolumn{1}{c|}{-128} \\\hline
\end{tabular}
\vskip0.2cm
\begin{tabular}{|c||c|c|c|} \hline
 $M^2_{D\alpha\beta}$
& \multicolumn{1}{c|}{\scriptsize{${\beta=1}$}} 
& \multicolumn{1}{c|}{\scriptsize{$\beta=2$}} 
& \multicolumn{1}{c|}{\scriptsize{$\beta=3$}} \\\hline\hline
 \scriptsize{$\alpha=1$}
& \multicolumn{1}{c|}{$(505)^2$} 
& \multicolumn{1}{c|}{0} 
& \multicolumn{1}{c|}{0} \\\hline

 \scriptsize{$\alpha=2$}
& \multicolumn{1}{c|}{0} 
& \multicolumn{1}{c|}{$(505)^2$} 
& \multicolumn{1}{c|}{0} \\\hline

 \scriptsize{$\alpha=3$}
& \multicolumn{1}{c|}{0} 
& \multicolumn{1}{c|}{0} 
& \multicolumn{1}{c|}{$(501)^2$} \\\hline
\end{tabular}
\hspace{0.4cm}
\begin{tabular}{|c||c|c|c|} \hline
 $M^2_{U\alpha\beta}$
& \multicolumn{1}{c|}{\scriptsize{${\beta=1}$}} 
& \multicolumn{1}{c|}{\scriptsize{$\beta=2$}} 
& \multicolumn{1}{c|}{\scriptsize{$\beta=3$}} \\\hline\hline
 \scriptsize{$\alpha=1$}
& \multicolumn{1}{c|}{$(508)^2$} 
& \multicolumn{1}{c|}{0} 
& \multicolumn{1}{c|}{0} \\\hline

 \scriptsize{$\alpha=2$}
& \multicolumn{1}{c|}{0} 
& \multicolumn{1}{c|}{$(508)^2$} 
& \multicolumn{1}{c|}{$(280)^2$} \\\hline

 \scriptsize{$\alpha=3$}
& \multicolumn{1}{c|}{0} 
& \multicolumn{1}{c|}{$(280)^2$} 
& \multicolumn{1}{c|}{$(387)^2$} \\\hline
\end{tabular}
\vskip0.4cm
\caption{\label{tab4}
The MSSM parameters at the scale Q=1 TeV in 
the QFV scenario based on the SPS1a' scenario. $T_{U \a\b}$ and $T_{D \a\b}$ are set 
to zero for $\a \neq \b$. All mass parameters are given in GeV. 
Note that $M^2_{U23}$=0 in the original SPS1a' scenario.
}
\end{center}
\end{table}

\begin{table}[t]
\begin{center}

\begin{tabular}{|c|c|c|c|c|c|} \hline
 
  \multicolumn{1}{|c|}{$\tilde u_1$} 
& \multicolumn{1}{c|}{$\tilde u_2$} 
& \multicolumn{1}{c|}{$\tilde u_3$} 
& \multicolumn{1}{c|}{$\tilde u_4$} 
& \multicolumn{1}{c|}{$\tilde u_5$} 
& \multicolumn{1}{c|}{$\tilde u_6$} \\\hline\hline
 
  \multicolumn{1}{|c|}{332} 
& \multicolumn{1}{c|}{541} 
& \multicolumn{1}{c|}{548} 
& \multicolumn{1}{c|}{565} 
& \multicolumn{1}{c|}{565} 
& \multicolumn{1}{c|}{612} \\\hline

\end{tabular}
\begin{tabular}{|c|c|c|c|c|c|} \hline
 
  \multicolumn{1}{|c|}{$\tilde d_1$} 
& \multicolumn{1}{c|}{$\tilde d_2$} 
& \multicolumn{1}{c|}{$\tilde d_3$} 
& \multicolumn{1}{c|}{$\tilde d_4$} 
& \multicolumn{1}{c|}{$\tilde d_5$} 
& \multicolumn{1}{c|}{$\tilde d_6$} \\\hline\hline
 
  \multicolumn{1}{|c|}{506} 
& \multicolumn{1}{c|}{547} 
& \multicolumn{1}{c|}{547} 
& \multicolumn{1}{c|}{547} 
& \multicolumn{1}{c|}{571} 
& \multicolumn{1}{c|}{571} \\\hline

\end{tabular}
\vskip0.2cm
\begin{tabular}{|c||c|c|c|c||c|c|} \hline
 
  \multicolumn{1}{|c||}{$\sg$} 
& \multicolumn{1}{c|}{$\tilde \chi^0_1$} 
& \multicolumn{1}{c|}{$\tilde \chi^0_2$} 
& \multicolumn{1}{c|}{$\tilde \chi^0_3$} 
& \multicolumn{1}{c||}{$\tilde \chi^0_4$} 
& \multicolumn{1}{c|}{$\tilde \chi^\pm_1$} 
& \multicolumn{1}{c|}{$\tilde \chi^\pm_2$} \\\hline\hline

  \multicolumn{1}{|c||}{608} 
& \multicolumn{1}{c|}{98} 
& \multicolumn{1}{c|}{184} 
& \multicolumn{1}{c|}{402} 
& \multicolumn{1}{c||}{415} 
& \multicolumn{1}{c|}{184} 
& \multicolumn{1}{c|}{417} \\\hline

\end{tabular}
\begin{tabular}{|c|c|c|c|} \hline
 
  \multicolumn{1}{|c|}{$h^0$} 
& \multicolumn{1}{c|}{$H^0$} 
& \multicolumn{1}{c|}{$A^0$} 
& \multicolumn{1}{c|}{$H^\pm$} \\\hline\hline
 
  \multicolumn{1}{|c|}{112} 
& \multicolumn{1}{c|}{426} 
& \multicolumn{1}{c|}{426} 
& \multicolumn{1}{c|}{434} \\\hline

\end{tabular}
\vskip0.4cm
\caption{\label{tab5}
Sparticles, Higgs bosons and corresponding physical masses (in GeV) in the 
scenario of Table \ref{tab4}. 
%
}
\end{center}
\end{table}

\begin{table}[t]
\begin{center}
\begin{tabular}{|c||c|c|c|c|c|c|}
		 \hline
			$|R^{\tilde u}_{i\a}|$
		  & $\ti u_L$ & $\ti c_L$ & $\ti t_L$ & $\ti u_R$ & $\ti c_R$ & $\ti t_R$ \\
		 \hline\hline
		  $\ti u_1$  & 0.010 & 0.032 & 0.457 & 0 & 0.369 & 0.809 \\
		  $\ti u_2$  & 0.014 & 0.015 & 0.691 & 0 & 0.720 & 0.062 \\
		  $\ti u_3$  & 0 & 0 & 0 & 1.0 & 0 & 0 \\
		  $\ti u_4$  & 0.896 & 0.444 & 0.011 & 0 & 0.003 & 0.001 \\
		  $\ti u_5$  & 0.443 & 0.893 & 0.036 & 0 & 0.062 & 0.008 \\
		  $\ti u_6$  & 0.021 & 0.058 & 0.559 & 0 & 0.585 & 0.585 \\

		 \hline
\end{tabular} 
		 
\vspace{3mm} 
%
%

\caption{\label{tab6} 
The up-type squark compositions in the flavour eigenstates, 
i.e. the absolute values of the mixing matrix elements $R^{\ti u}_{i\a}$, 
at the scale Q=1 TeV for the scenario of Table \ref{tab4}.
}

\end{center}
\end{table}
%

In Fig.5 we show the $\delta^{uRR}_{23}$ 
dependence of the QFV production cross section $\sigma^{11}_{ct}$ at $E_{cm}$ = 7 TeV
and 14 TeV, where all basic parameters other than $M^2_{U 23}(Q=1 TeV)$ are fixed as 
in the scenario of Table \ref{tab4}.  $\sigma^{22}_{ct}$ is very small due to the very small 
$B(\su_2 \to t \nt_1)$. 
We see that the QFV cross section increases with increase of the QFV parameter 
$|\delta^{uRR}_{23}|$ and that it can be quite sizable in a wide allowed range of 
$\delta^{uRR}_{23}$. The mass of $\su_1$ decreases and $B(\su_1 \to c \nt_1) \cdot 
B(\su_1 \to t \nt_1)$ increases with increase of $|\delta^{uRR}_{23}|$. 
This leads to the increase of $\sigma^{11}_{ct}$ with increase of $|\delta^{uRR}_{23}|$.
$\sigma^{11}_{ct}$ vanishes for $|\delta^{uRR}_{23}| \gsim 0.62$, where the decay 
$\su_1 \to t \nt_1$ is kinematically forbidden.
%
\begin{figure}
\begin{center}
\scalebox{0.6}[0.6]{\includegraphics{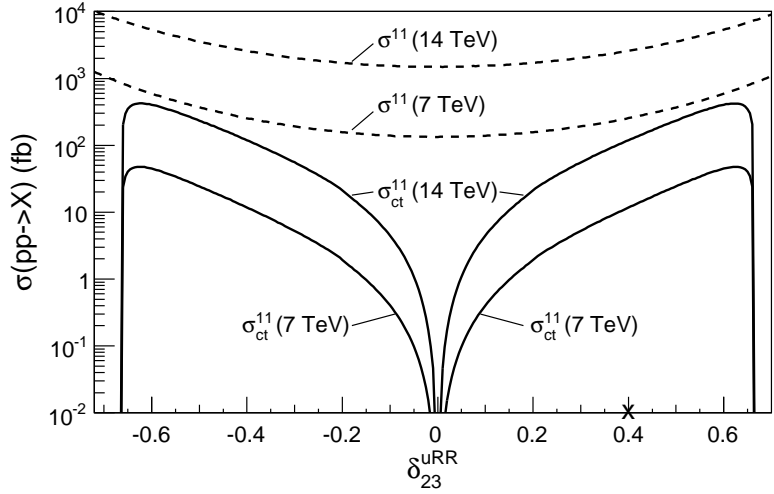}}
\caption{
$\delta^{uRR}_{23}$ dependences of $\sigma^{11} \equiv \sigma(pp \to \su_1 \bar{\su}_1 X)$ and 
$\sigma^{11}_{ct}$ at $E_{cm}$ = 7 TeV and 14 TeV. 
The point "x" of $\delta^{uRR}_{23} = 0.4$ corresponds to the QFV scenario of Table 
\ref{tab4}. The shown range of $\delta^{uRR}_{23}$ is allowed by the conditions (i) to 
(v) given in the text. 
%
}
\label{fig5}
\end{center}
\end{figure}
%
%

Finally, we briefly discuss the detectability of the QFV 
production process $pp \to \su_i \bar{\su}_i X \to c \bar{t} (t \bar{c}) 
\nt_1 \nt_1 X$ $(i=1,2)$ at LHC. The signature is 
'(anti)top-quark + charm-jet + $E_T^{mis}$ + $X$', where X contains beam-jets only. 
Therefore, identifying the top-quarks in the final states is mandatory. 
This should be possible by using the hadronic decays of the top-quark. 
Charm-tagging would also be very useful.
%
There is another QFV signal process leading to the same final states, i.e. 
gluino production and its QFV decay \cite{Bartl:QFV_gluino_decay} 
$pp \to \sg \nt_1 X \to c \bar{t} (t \bar{c}) \nt_1 \nt_1 X$. This cross section 
is, however, about a factor of 20-30 smaller than that of the QFV process via 
squark pair production in the scenarios studied here. This suppression is mainly 
due to the electroweak interactions involved. 
The QFV production process $pp \to c \bar{t} (t \bar{c}) X$ via SUSY-QCD one-loop diagrams 
\cite{Hollik: single top} also yields the signature '(anti)top-quark + charm-jet + X'. 
The size of the cross section of this process is of the same order as 
that of our QFV process Eq.(\ref{eq:QFVsq_production_decay}). 
However, the missing-$E_T$ would be much smaller than that in our signal process 
Eq. (\ref{eq:QFVsq_production_decay}). \\
%
%
If charm-tagging is not possible, one should search for the process 
$pp \to \su_i \bar{\su}_i X \to q \bar{t} (t \bar{q}) \nt_1 \nt_1 X$ 
($q \neq t, b$), i.e. for the signature '(anti)top-quark + jet + $E_T^{mis}$ + $X$'. 
Main backgrounds are single top-quark productions in the SM. 
The most important one is due to tW production where the W-boson 
decays into a tau-lepton which then decays hadronically: 
$pp \to t W X \to t \tau \nu X \to t \, \tau$-jet $\nu \bar{\nu} X$.
The cross section for the tW production $\sigma(pp \to t W^- X) + 
\sigma(pp \to \bar{t} W^+ X)$ is about 66 pb 
for $E_{cm}$=14 TeV \cite{Beneke:2000hk}. 
It turns out that the W-boson is mainly produced in the central 
region (see e.g.~\cite{Re:2010bp}). 
To reduce this background one can use the fact that a charm-quark jet 
has usually a much higher particle-multiplicity 
than the $\tau$-jet. By requiring that at least four hadrons are 
contained in the jet, this background cross section can be reduced 
to about 12 fb for $E_{cm}$= 14 TeV without significant loss of 
our QFV signal events. 
One can expect that it is much smaller than 12 fb for $E_{cm}$= 7 TeV.
%
%
In the case that one considers only hadronic decays of the top-quark, 
one can require in addition that the invariant mass of each jet 
is larger than the tau-lepton mass, which should reduce this background 
further again without significant loss of our signal events. \\
%
%
%
%
%
The second important background is single top-quark production due to 
t-channel W-boson exchange in the SM. Relevant for us is the reaction 
$pp \to t(\bar{t}) q(\bar{q}) Z^0 X  \to t(\bar{t}) q(\bar{q}) \nu \bar{\nu} X$, 
where the main contribution is due to W-boson exchange in the t-channel 
(quite similar to $pp \to t \, q\, X$, for which a thorough treatment is 
given in \cite{1-3 mixing}). 
Using the WHIZARD/O'MEGA package \cite{Whizard, Omega} we have calculated the 
corresponding cross sections and obtained at $E_{cm}$ = 14 TeV [7 TeV]: \\
~~$\sigma(pp \to t q Z^0 X  \to t q \nu \bar{\nu} X)$ = 97.4 [17.5] fb, \\
~~$\sigma(pp \to t \bar{q} Z^0 X  \to t \bar{q} \nu \bar{\nu} X)$ = 15.9 [1.89] fb, \\
~~$\sigma(pp \to \bar{t} q Z^0 X  \to \bar{t} q \nu \bar{\nu} X)$ = 46.0 [7.1] fb, \\
~~$\sigma(pp \to \bar{t} \bar{q} Z^0 X  \to \bar{t} \bar{q} \nu \bar{\nu} X)$ = 13.2 [1.6] fb, \\
where the cross sections summed over $q = u,d,c,s$ are shown. 
Comparing these to Figs. 4 and 5 we see that for $|\delta_{23}^{uRR}| \gsim 0.45$ 
our signal cross section is larger than the sum of these background cross sections. 
For $|\delta_{23}^{uRR}| \lsim 0.45$ a suitable $E_T^{mis}$ cut will reduce this background 
relatively to our signal because the $E_T^{mis}$ due to $\nu \bar{\nu}$ from the $Z^0$ 
decay will on the average be smaller than that due to the two neutralinos. \\
%
The cross section of single top-quark production via s-channel W-boson exchange 
is much smaller than that of our QFV process. We obtain for $E_{cm}$ = 14 TeV [7 TeV]: \\
~~$\sigma(pp \to "W^+" Z^0 X  \to t \bar{b} \nu \bar{\nu} X)$ = 1.42 [0.45] fb, \\
~~$\sigma(pp \to "W^-" Z^0 X  \to \bar{t} b \nu \bar{\nu} X)$ = 0.73 [0.18] fb. \\
There could be another background from the QFC top-quark pair production processes 
(a) $pp \to \su_i \bar{\su}_i X \to t \bar{t} \nt_1 \nt_1 X$ and 
(b) $pp \to t \bar{t} Z^0 X \to t \bar{t} \nu \bar{\nu} X$, where one of the 
W-bosons from the top-quarks decays leptonically with the charged lepton 
being missed. However, the probability of such W-boson decay would be 
very small. Moreover, these top-quark pair production cross sections are not so large 
compared to our QFV cross sections. For the process (a), for example, we see that 
$\sigma^{ii}_{t\bar{t}} < \sigma^{ii}_{ct}$  $(i=1,2)$ in the scenario studied 
as shown just after Eq. (\ref{eq:QFC_squark_production_decay}). 
For (b) we obtain at $E_{cm}$ = 14 TeV [7 TeV]: 
$\sigma(pp \to t \bar{t} Z^0 X \to t \bar{t} \nu \bar{\nu} X)$ = 97.6 [13.8] fb. 

Of course, a detailed Monte Carlo study including detector effects is 
required for a proper assessment of the detectability of the  
proposed QFV signal. 
However, this is beyond the scope of this paper and will be presented 
in a forthcoming publication \cite{newpaper}. 

%

\section{Conclusion \label{Conclusion}}

To conclude, we have studied the effects of squark mixing of the second and 
third generation, especially $\ti c_{L/R}$ - $\ti t_{L/R}$ mixing, on squark 
production and decays at LHC in the MSSM. We have shown that the effect 
can be very large in a significant region of the QFV parameters despite the very 
strong constraints on QFV from experimental data on B mesons. 
The QFV squark decay branching ratios B($\su_i \to c \nt_1$) and B($\su_i \to t \nt_1$) 
$(i=1,2)$ can be very large (up to $\sim$ 50\%) simultaneously.
This can result in QFV signal events '$pp \to$ $c$${\bar t}$ ($t$${\bar c}$) + 
$E_T^{mis}$ + beam-jets' with a significant rate at LHC. 
The observation of these remarkable signatures would provide a 
powerful test of supersymmetric QFV at LHC. 
Therefore, in the squark search one should take into account the possibility of 
significant contributions from QFV squark decays. 
Moreover, one should also include the QFV squark parameters (i.e. the squark 
generation mixing parameters) in the determination of the basic SUSY 
parameters at LHC.


\section*{Acknowledgments}

This work is supported by the "Fonds zur F\"orderung der
wissenschaftlichen Forschung (FWF)" of Austria, project No. P18959-N16.
The authors acknowledge support from EU under the MRTN-CT-2006-035505
network program. 
B. H. and  W. P. are supported by the DFG, project No. PO 1337/1-1.
The work of B. H. is supported in part by the Landes-Exzellenzinitiative Hamburg.
W. P. is supported by the Alexander von Humboldt Foundation and
the Spanish grant FPA2008-00319/FPA.



\end{document}